\documentclass[journal]{IEEEtran}  

\usepackage[]{graphicx,float,latexsym,amssymb,amsfonts,amsmath,amstext,times}
\usepackage{cite}
\newcommand{\ignore}[1]{}  
\usepackage{amsmath}
\usepackage[dvipsnames]{xcolor}
\usepackage{boxhandler}
\usepackage{siunitx}
\usepackage{balance}
\usepackage{cleveref}
  \usepackage{times}
\usepackage[T1]{fontenc}
\usepackage{mwe}    
\usepackage{subfigure}
\usepackage {optidef}
\usepackage{tablefootnote}
\usepackage{footnote}
\makesavenoteenv{tabular}
\usepackage[para,online,flushleft]{threeparttable}
\usepackage{enumitem}

\usepackage{booktabs, tabularx}

\usepackage[hyphens]{url}
\usepackage[]{graphicx}    
\usepackage[T1]{fontenc}
\usepackage{array}
\newcolumntype{P}[1]{>{\centering\arraybackslash}p{#1}}
\usepackage{multirow}
\usepackage{multicol}
\usepackage{mathtools}
\usepackage{graphicx}
\usepackage{balance}
\renewcommand{\baselinestretch}{0.95}
\begin{document}
\title{Detection and Classification of UAVs Using \\RF Fingerprints in the Presence of Interference
}

\author{%
Martins Ezuma, Fatih Erden, Chethan Kumar Anjinappa, Ozgur Ozdemir, \IEEEmembership{Member,~IEEE,} and\\ Ismail Guvenc,~\IEEEmembership{Senior Member,~IEEE}%
\vspace{-5mm}
\thanks{This work has been supported in part by NASA under the Federal Award ID number NNX17AJ94A. This paper was presented in part at the IEEE Aerospace Conference, Big Sky, Montana, Mar. 2019~\cite{ezuma2019micro}. First three authors of this paper have equal contributions.}
\thanks{All the authors are with the Department of Electrical and Computer Engineering, North Carolina State University, Raleigh, NC 27606 (e-mail:
\{mcezuma, ferden, canjina, oozdemi, iguvenc\}@ncsu.edu).}%
}
\maketitle
\thispagestyle{plain}
\pagestyle{plain}

\begin{abstract}
This paper investigates the problem of detection and classification of unmanned aerial vehicles (UAVs) in the presence of wireless interference signals using a passive radio frequency (RF) surveillance system. The system uses a multistage detector to distinguish signals transmitted by a UAV controller from the background noise and interference signals. First, RF signals from any source are detected using a Markov models-based na\"ive Bayes decision mechanism. When the receiver operates at a signal-to-noise ratio (SNR) of 10~dB, and the threshold, which defines the states of the models, is set at a level 3.5 times the standard deviation of the preprocessed noise data, a detection accuracy of 99.8\% with a false alarm rate of 2.8\% is achieved. Second, signals from Wi-Fi and Bluetooth emitters, if present, are detected based on the bandwidth and modulation features of the detected RF signal. Once the input signal is identified as a UAV controller signal, it is classified using machine learning (ML) techniques. Fifteen statistical features extracted from the energy transients of the UAV controller signals are fed to neighborhood component analysis (NCA), and the three most significant features are selected. The performance of the NCA and five different ML classifiers are studied for 15 different types of UAV controllers. A classification accuracy of 98.13\% is achieved by k-nearest neighbor classifier at 25~dB SNR. Classification performance is also investigated at different SNR levels and for a set of 17 UAV controllers which includes two pairs from the same UAV controller models.


\end{abstract}



\section{Introduction}


\IEEEPARstart{U}{nmanned} aerial vehicles (UAVs), or drones, are becoming ubiquitous in modern society. The recent popularity of UAVs is mainly due to the advancement in micro-electro-mechanical systems-based precision sensors, such as inertial motion units and gyroscopes, which are used for guidance, navigation, and control of UAVs. Consequently, UAVs have become relatively cheap and affordable. They are finding new applications in areas such as surveillance, smart policing, search and rescue missions, infrastructure inspections, package delivery, and precision agriculture\cite{shakhatreh2018unmanned}.
Judging by the current trend in UAV applications, it is expected that UAVs will become an integral part of modern society. However, there are security and privacy issues associated with the ubiquity of UAVs.

In recent times, UAVs have been used in ways that introduce a threat to public safety~\cite{guvencc2017detection}. There have been several instances where hobby drones have been used to transport illegal drugs across prison walls. In addition, drones have carried out espionage attacks which pose serious risk to public safety. Recently, drones operated by dissidents have flown into sensitive national infrastructures like nuclear reactors and airports~\cite{solodov2018analyzing}. Moreover, drones are becoming tools for cyberattack and terrorism. For instance, Wi-Fi sniffing UAVs can eavesdrop on smartphone users and steal sensitive data without being detected~\cite{alajmi2017uavs,nassi2019sok}.

Considering the security and privacy issues associated with UAVs, accurate detection and classification of these vehicles are vital to public safety and national security. One promising technique for UAV detection is based on the analysis of radio frequency (RF) signals from UAV controllers. In~\cite{shoufan2018drone}, drone pilots are identified by analyzing RF signals captured from the drone controllers. The pilots' behavioral biometrics can be identified from the captured signals using machine learning (ML) techniques. The ML algorithms are trained using the RF signals when the controller is handled by the legitimate owner of the device. That way, it is possible to identify different drones and their pilots. However, while the behavioral biometrics of drone pilot is an important information for drone detection, an adversary could be anyone whose behavior metrics we have no prior knowledge of. Therefore, in order to accurately detect and identify an adversary drone, one should focus on identifying the intrinsic signature of the drone controller itself. These intrinsic signatures can be extracted from the RF signals transmitted by the UAV controllers and referred to as the \textit{RF fingerprints} of the controllers.

Since the communication signals of most commercial and hobby grade UAVs are transmitted in the same frequency band as Wi-Fi and Bluetooth transmissions, it becomes challenging to detect and identify RF signals from the UAV controllers in the presence of these interferers. Moreover, surveillance and electronic warfare systems should be able to differentiate UAVs from different manufacturers. For instance, the identity of a UAV can provide useful information about the payload, operational range, control signal characteristics (e.g., for jamming such signals), and the threat capability of the associated UAV. Accurate identification of UAVs is also important in digital forensic analysis of aerial threats.

In this work, we propose a multistage UAV detection and an ML-based classification system for identifying 17 different UAV controllers in the presence of wireless interference, i.e., Wi-Fi and Bluetooth devices. The multistage UAV detection system consists of two detectors. The first detector employs a two-state Markov model based na\"ive Bayes algorithm in deciding if the captured data contains RF signals or not. Once an RF signal is detected, the second stage detector decides if the signal comes from a UAV controller or an interference source. Given that the detected RF signal is from an interference source, the source class is identified as Wi-Fi or Bluetooth. On the other hand, if the detected signal is from a UAV controller, the signal is transferred to the ML-based classification system to determine the make and model of the UAV controller. In an earlier work~\cite{ezuma2019micro}, the authors proposed a system for detecting and classifying 14 different UAV controllers. The system design assumes the absence of interference signals. However, this assumption is not always correct. The contributions of the current work are summarized as follows:\looseness=-1


\begin{enumerate}
\item The paper investigates the problem of detecting and classifying signals from UAV controllers in the presence of co-channel wireless interference. We consider interference from Wi-Fi and Bluetooth sources and describe a methodology to detect the UAVs. The interference detection ensures the proposed UAV detection system is robust against false alarms and missed target detection. In addition, in~\cite{ezuma2019micro}, we used two fixed thresholds, positioned at $\pm$ 3$\sigma$, to transform the captured signal into three-state Markov models, where $\sigma$ is the standard deviation of the noise signal in the environment. However, in the current work, we use a single threshold to transform the captured signal into two-state Markov models, which reduces overall complexity. We also define a procedure to determine the optimum threshold value based on the available training data.
Besides, in the current study, we evaluate the detection performance for different thresholds based on the false alarm rate (FAR).

\item We introduce the concept of \textit{energy transient} for the extraction of RF-based features and show how effective it is for the classification of the UAV controller signals. The energy transient is computed using the representation of the RF signals in energy-time-frequency domain. From the energy transient, 15 statistical features are extracted for the UAV classification. The performance of five different ML algorithms are compared using the proposed RF fingerprinting technique. In addition, we investigate the neighborhood component analysis (NCA) as a practical algorithm for feature selection in the classification problem. The classification results using the three most significant features, selected by the NCA, are compared with those when all the 15 RF features are used. We also evaluate the classification performance at different signal-to-noise ratios (SNRs). For an SNR of 25 dB, the results show that the k-nearest neighbor (kNN) and random forest (RandF) are the best performing classifiers, achieving accuracy of 98.13\% and 97.73\%, respectively, when the three most significant RF-based features are used for the classification of 15 UAV contollers.

\item We study the confusion that results when attempting to classify UAV controllers of the same make and model. This is important in digital forensic analysis and detecting decoys in surveillance systems. To investigate this confusion, we included two pairs of identical UAV controllers to a pool of 13 different UAV controllers. That is, we capture control signals from 17 UAV controllers and evaluate the ability of the proposed classification system at different SNRs. For an SNR of 25 dB, kNN and RandF achieve accuracy of 95.53\% and 95.18\%, respectively, when the three most significant RF features are used. To the best of our knowledge, past studies on UAV classification using RF techniques considers only a limited number of different make and model UAV controllers, often less than 10~\cite{Caidanhash, bisio2018unauthorized}.
\end{enumerate}


The remainder of the paper is organized as follows:  Section~\ref{two} provides a brief overview of the related work. Section~\ref{two_1} describes the multistage detection system, and Section~\ref{second_stage_detector} introduces the methodology to detect Wi-Fi and Bluetooth interference signals. Feature extraction and the RF fingerprinting-based UAV classification system are explained in Section~\ref{four_1}. The experimental setup is described in Section~\ref{four} while the detection and classification results are presented in Section~\ref{results_1}. The paper is concluded in Section~\ref{label:conclusion}.

\section{Related Work} \label{two}



UAV detection and classification through RF signals can be grouped into two major headings: RF fingerprinting and Wi-Fi fingerprinting techniques. These techniques use an RF sensing device to capture the RF communication signal between a UAV and its controller. In RF fingerprinting techniques, physical layer features and signatures of the captured signal are used for the detection and classification of the UAV or its controller. On the other hand, Wi-Fi fingerprinting techniques extracts the medium access control (MAC) and network layer features of the captured UAV RF transmission. \looseness=-1

  \begin{figure}[t!]
    \centering
    \includegraphics[width=\linewidth]{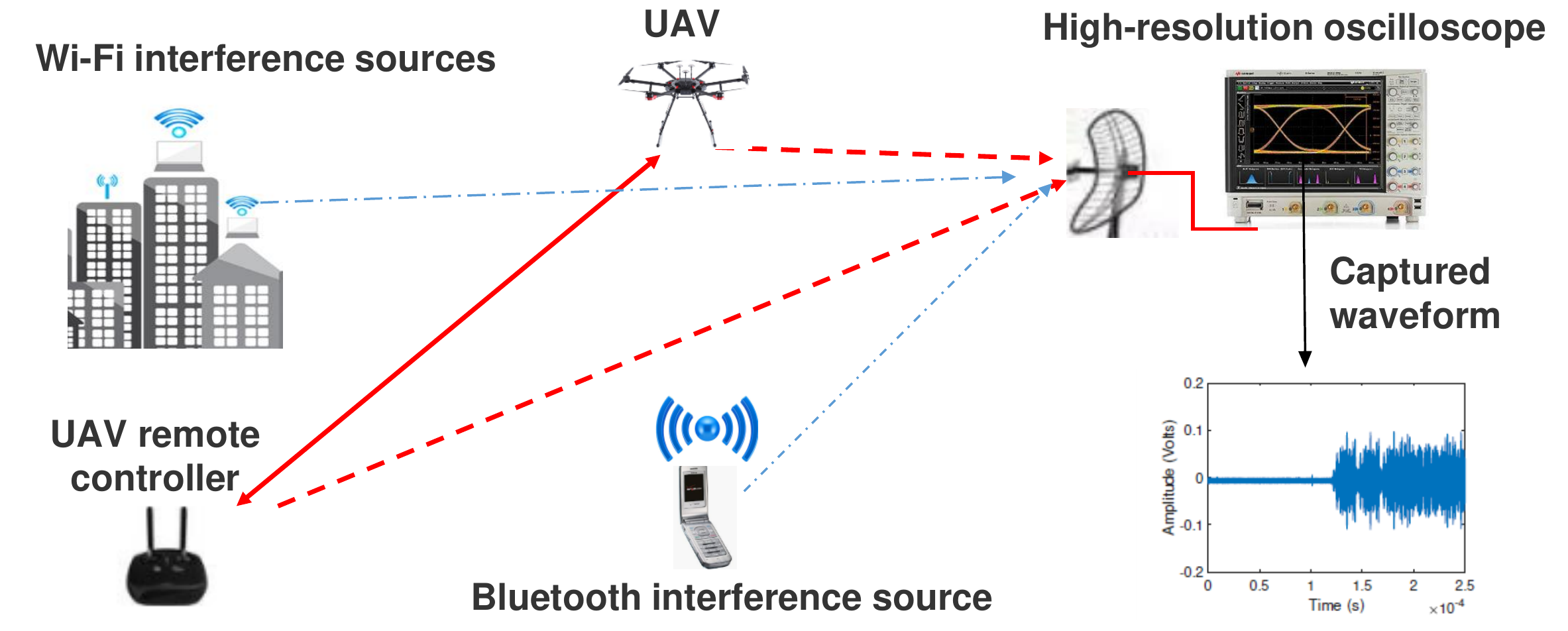}
    \caption{The scenario of the RF-based UAV detection system.}
    \label{RFSystem_setup}
    \vspace{-3mm}
\end{figure}

\begin{figure*}[t!]
\center{
\begin{subfigure}[]{\includegraphics[scale=0.28]{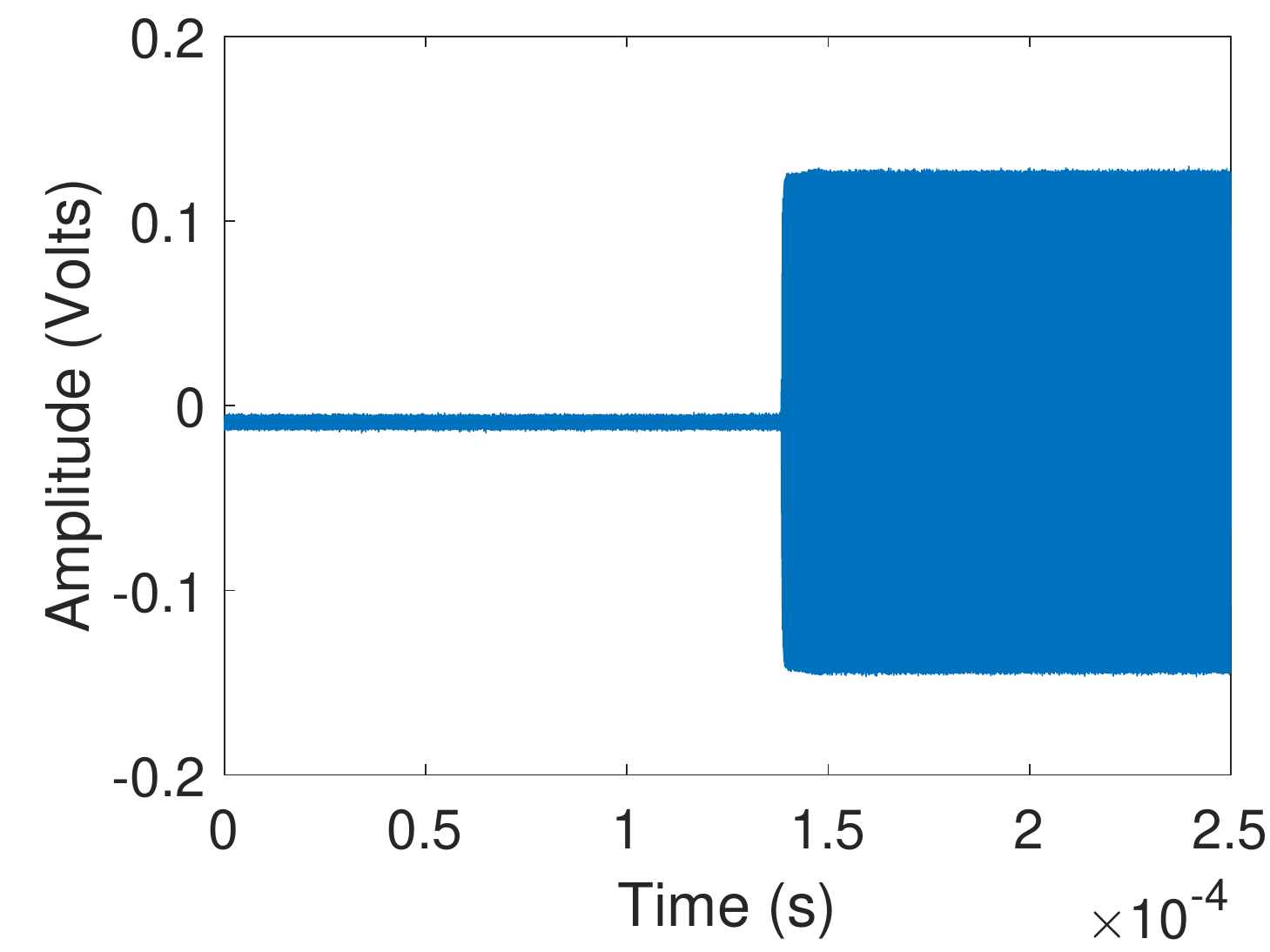}\label{Graupner_MC-32}}
\end{subfigure}
\begin{subfigure}[]{\includegraphics[scale=0.28]{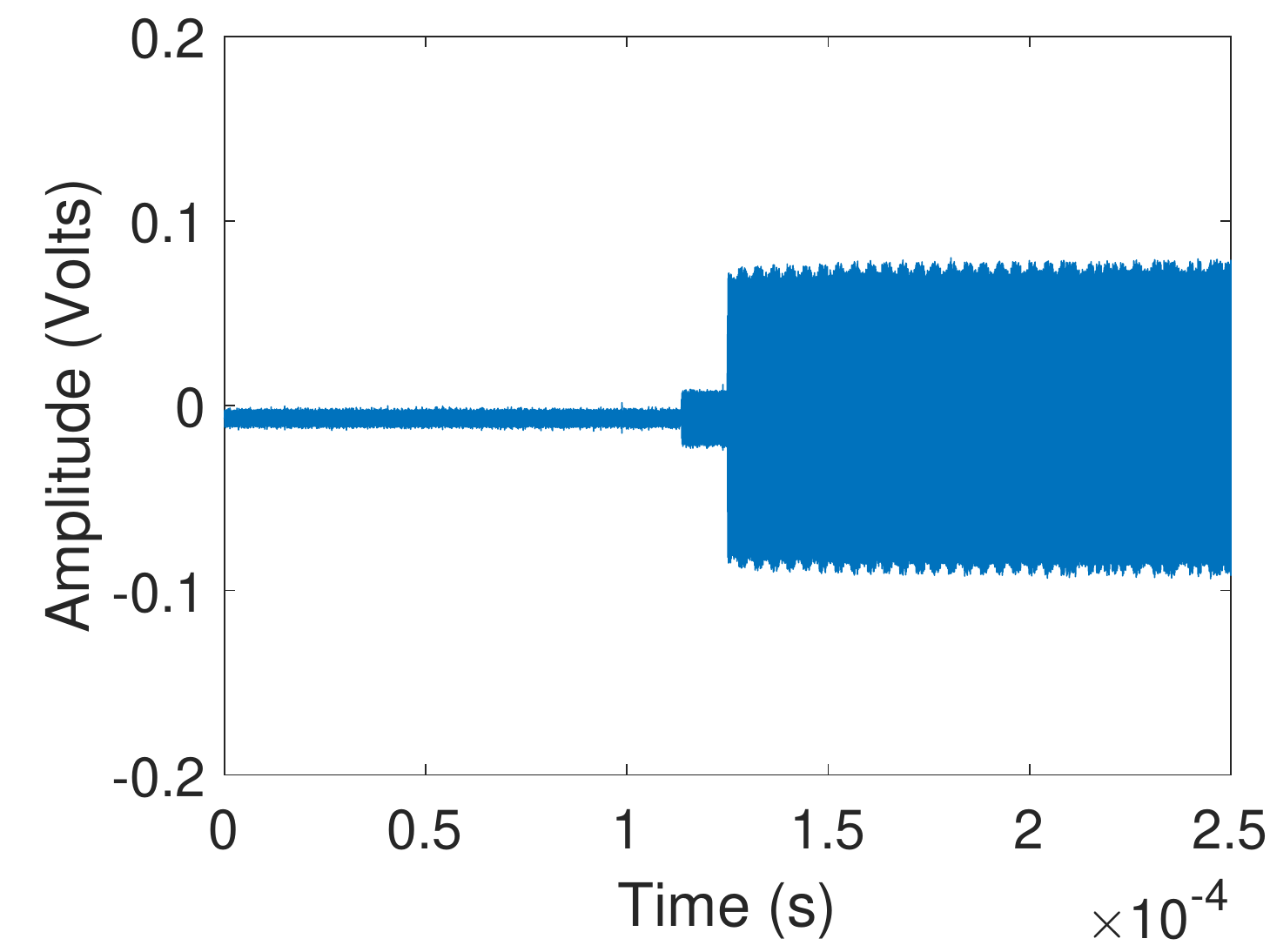}\label{Raw_data_DX6e}}
\end{subfigure}
\begin{subfigure}[]{\includegraphics[scale=0.28]{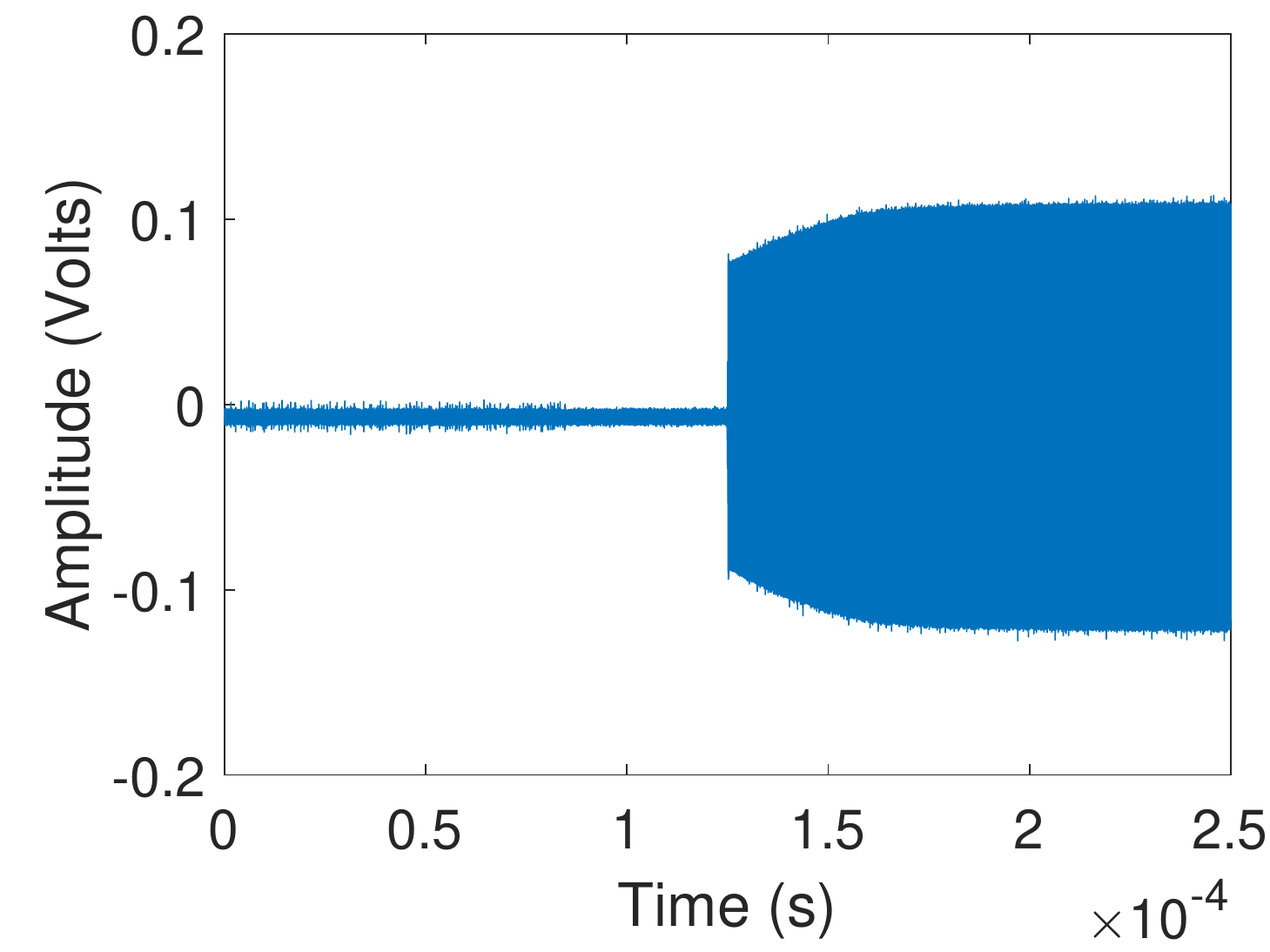}\label{T14SG}}
\end{subfigure}
\begin{subfigure}[]{\includegraphics[scale=0.28]{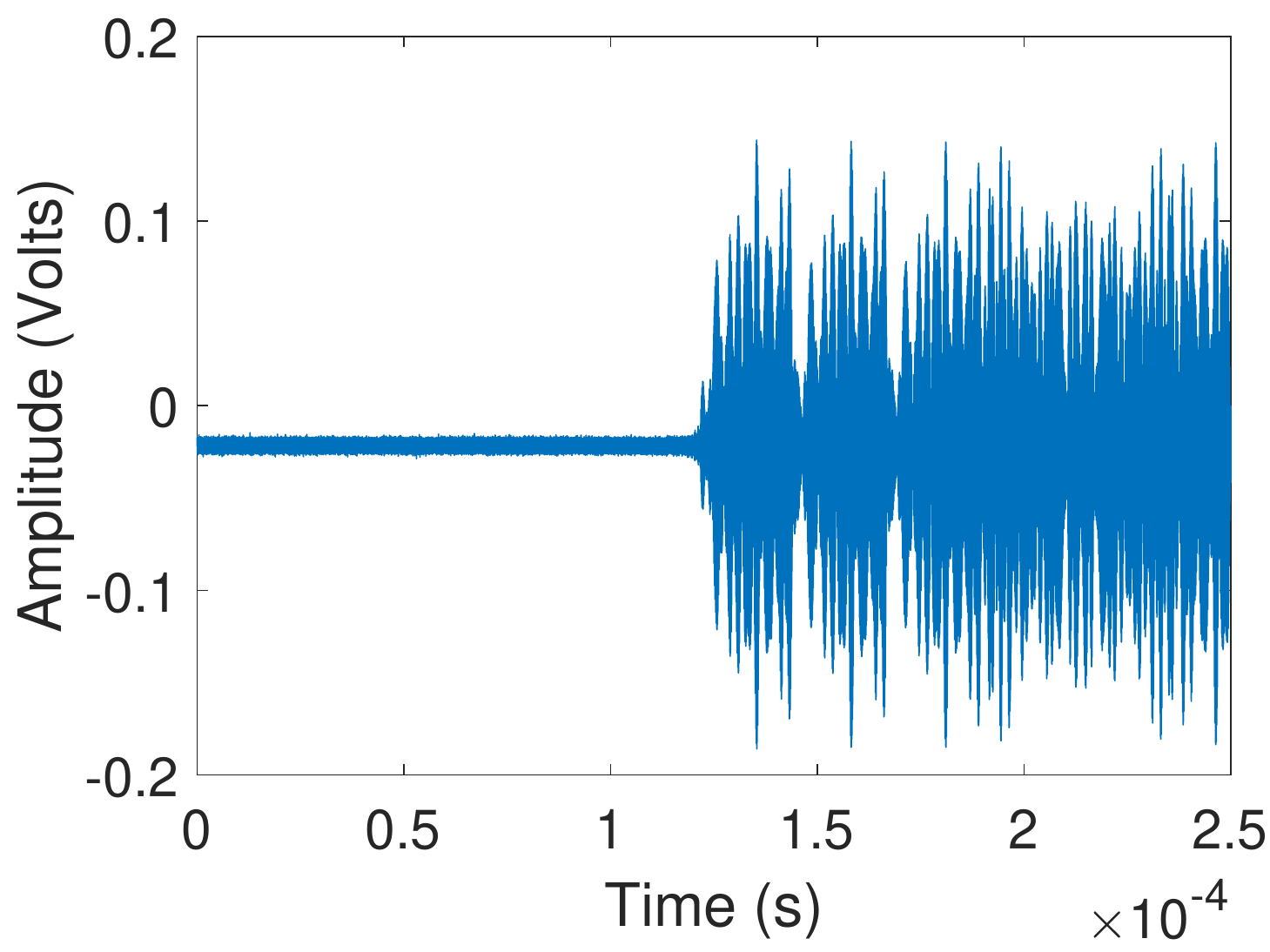}\label{DJI_Phantom4Pro}}
\end{subfigure}
\caption{{RF signals captured from four different UAV controllers: (a) Graupner MC-32, (b) Spektrum DX6e, (c) Futaba T8FG, and (d) DJI Phantom 4 Pro.}}
 \label{fig:UAV data waveforms}}
 \vspace{-3mm}
 \end{figure*}

\subsection{RF Fingerprinting-Based UAV Detection and Classification}
RF fingerprinting techniques rely on the unique characteristics of the RF signal waveforms captured from different UAV controllers. Experimental investigations show that most of the commercial UAVs have unique RF signatures which is due to the circuitry design and modulation techniques employed. Therefore, RF fingerprints extracted from the UAV or its remote controller signals can be used as a basis for detection and classification of the UAVs.

In~\cite{zhao2018classification}, RF fingerprints of the UAV's wireless control signals are extracted by computing the amplitude envelope of the signal. The dimensionality of the processed signal is reduced by performing principal component analysis (PCA), and the lower-dimensional data is fed into an auxiliary classifier Wasserstein
generative adversarial networks (AC-WGANs). The AC-WGANs achieves an overall classification rate of 95\% when four different types of UAVs are considered. \looseness=-1

In~\cite{al2019rf}, drones are detected by analyzing the RF background activities along with the RF signals emitted when the drones are operated in different modes. Afterward, RF spectrum of the drone signal is computed using the discrete Fourier transform (DFT). The drone classification system is designed by training a deep neural network with the RF spectrum data of different drones. The system shows an accuracy of 99.7\% when two drones are classified, 84.5\% with four drones, and 46.8\% with ten drones.

In~\cite{boon2017rf}, an industry integrated counter-drone solution is described. The solution is based on a network of distributed RF sensors. In this system, RF signals from different UAV controllers are detected using an energy detector. Afterward, the signals of interest are classified using RF spectral shape correlation features. Besides, distributed RF sensors make it possible to localize the UAV controller using time difference of arrival (TDoA) or multilateration techniques. However, this industrial solution is quite expensive.

\subsection{Wi-Fi Fingerprinting-Based UAV Detection and Classification}
Wi-Fi fingerprinting-based techniques are motivated by the fact that some UAVs use Wi-Fi links for control and video streaming. The RF sensing system consists primarily of a Wi-Fi packet-sniffing device, which can intercept the Wi-Fi data traffic between a UAV and its remote controller. In~\cite{bisio2018unauthorized}, unauthorized Wi-Fi controlled UAVs are detected by a patrolling drone using a set of Wi-Fi statistical features. The extracted features include MAC addresses, root-mean-square (RMS) of the Wi-Fi packet length, packet duration, average packet inter-arrival time, among others. These features are used to train different ML algorithms which perform the UAV classification task. In~\cite{bisio2018unauthorized}, the random tree and random forest classifiers achieve the best performance as measured by the true positive and false positive rates.

In~\cite{nguyen2017matthan}, drone presence is detected by eavesdropping on Wi-Fi channels between the drone and its controller. The system detects drones by analyzing the impact of their unique vibration and body shifting motions on the Wi-Fi signals transmitted by the drone. The system achieves accuracy above 90\% at 50 meters.

In general, a major concern with the Wi-Fi fingerprinting techniques is the privacy. This is because the same Wi-Fi detection system can spoof Wi-Fi traffic data from a smartphone user or a private Wi-Fi network. In addition, only a limited number of commercial drones employ Wi-Fi links for video streaming and control. Most commercial drones use proprietary communication links.

Besides RF and Wi-Fi fingerprinting techniques, several other techniques have been investigated for UAV detection, including radar-based techniques, acoustic techniques, and computer vision techniques~\cite{guvenc2018detection}. However, as discussed in~\cite{guvenc2018detection},  traditional radar systems are not so effective in detecting UAVs with small radar cross sections, and acoustic and computer vision-based techniques are greatly impaired by ambient environmental conditions. In contrast, RF techniques are not limited by these problems. We start by describing the design of the multistage detector of our proposed RF-based system.

\section{Multistage UAV Signal Detection}\label{two_1}

We consider the scenario shown in Fig.~\ref{RFSystem_setup}, where a passive RF surveillance system listens for the control signals transmitted between a UAV and its remote controller. The main hardware components of the surveillance system are 2.4~GHz RF antenna and a high-frequency oscilloscope, which is capable of sampling the captured data at 20 Gsa/s. The high sampling rate enables the surveillance system to capture high-resolution waveform transient features of any detected RF signal. The waveform transient features of different UAV controllers are unique. This is a useful property for detecting and classifying RF signals from different UAV controllers. Fig.~\ref{fig:UAV data waveforms} illustrates sample RF signals captured from four different UAV controllers. The figure shows each signal has a distinct waveform transient or shape which can be exploited for identifying the source UAV controller.

Since most commercial UAVs operate in the 2.4 GHz band, the passive RF surveillance system is designed to operate in this frequency band. However, this also corresponds to the operational band of Wi-Fi and mobile Bluetooth devices. Therefore, in real wireless environment, signals from these wireless sources will act as interference to the detection of the UAV control signals.

\begin{figure}[t!]
\center{\includegraphics[trim=0.1cm 0cm 0.1cm 0.5cm, clip,scale=0.5]{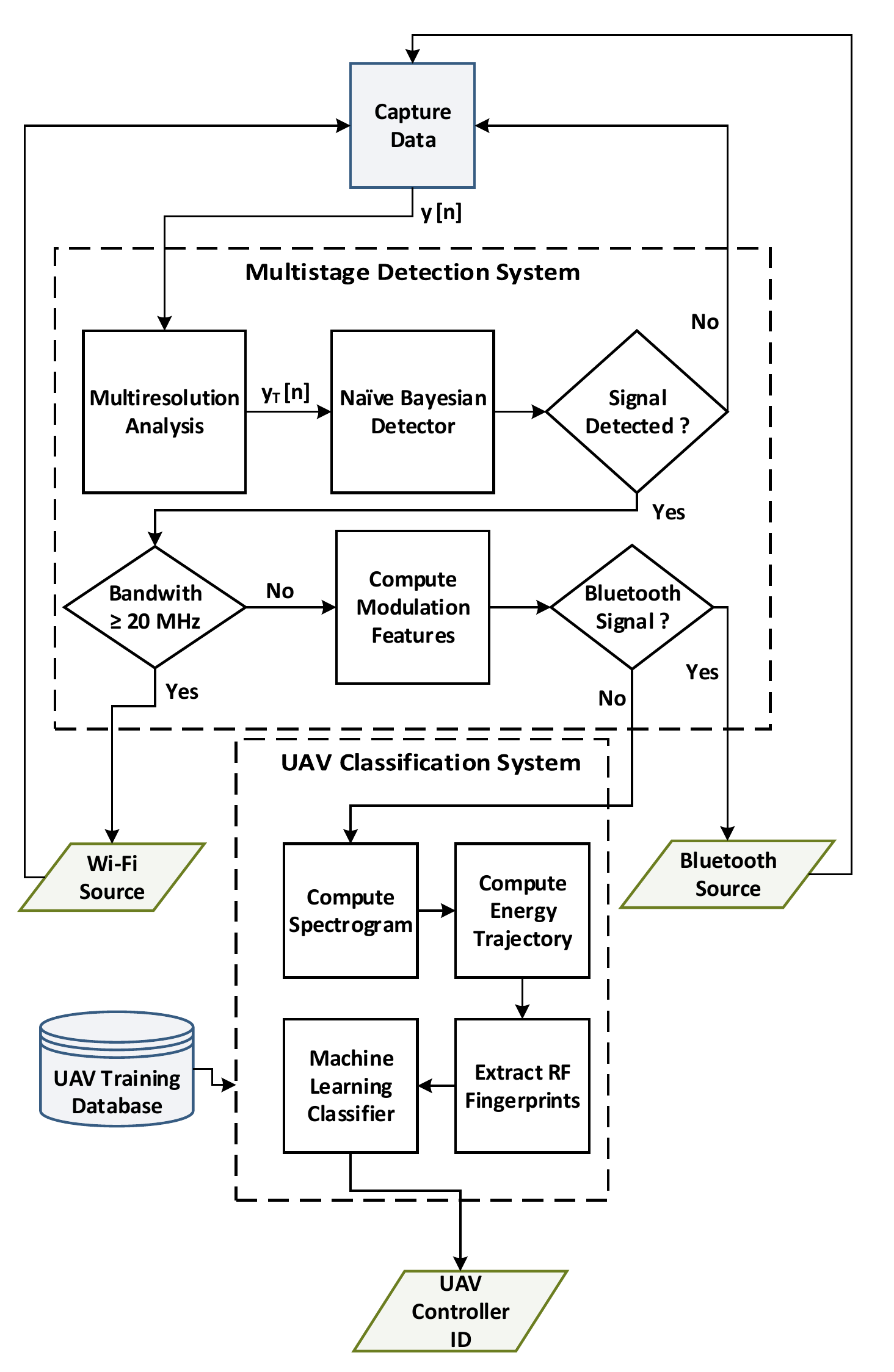}}
\caption{The system flowchart providing a graphical description of information processing and flow of data through the system.}
\label{Fig:flowchart}
\vspace{-5mm}
\end{figure}

Given the scenario in Fig.~\ref{RFSystem_setup}, the passive RF surveillance system has to decide if the captured data comes from a UAV controller, an interference source, or background noise. In the case where the captured data comes from a UAV controller, the detection system should be able to correctly classify the UAV controller. However, if the detected signal is from an interference source, the detection system should be able to correctly identify the source, i.e., a Wi-Fi or a Bluetooth device. Therefore, the detection problem is a multi-hypothesis problem. For such problems, it is well known that computational complexity increases as the number of hypothesis increases. Consequently, the multi-hypothesis detection problem can be simplified by using a multistage sequential detector. In this system, each detection stage is a simple binary hypothesis test which is much easier to solve. \looseness=-1

The flowchart in Fig.~\ref{Fig:flowchart} provides a high-level graphical description of the entire system. The first step in detecting and identifying a UAV controller is data capture. Usually, the captured raw signal has a large size and is often very noisy. Therefore, before detection and classification, the signals are first pre-processed using multiresolution analysis. Next, the processed signals are transferred to the multistage detection system, which consists of two stages. In the first stage, the detector employs na\"ive Bayesian hypothesis test in deciding if the captured signal is an RF signal or noise. If the decision is positive, the second stage detector is activated to decide if the captured RF signal comes from an interference source or a UAV controller. This detector uses bandwidth analysis and modulation-based features for interference detection. If the detected RF signal is not from a Wi-Fi or Bluetooth interference source, it is presumed to be a signal transmitted by a UAV controller. Consequently, the detected signal is transferred to an ML-based classification system for accurate identification of the UAV controller. \looseness=-1

\subsection{Pre-processing Step: Multiresolution Analysis}
Captured RF data are pre-processed by means of wavelet-based multiresolution analysis. It has been established that multiresolution decomposition using discrete wavelet transform (DWT) like the Haar wavelet transform is effective for analyzing the information content of signals and images~\cite{mallat1989theory}. \looseness=-1

\begin{figure}[t!]
    \centering
    \includegraphics[scale=0.31]{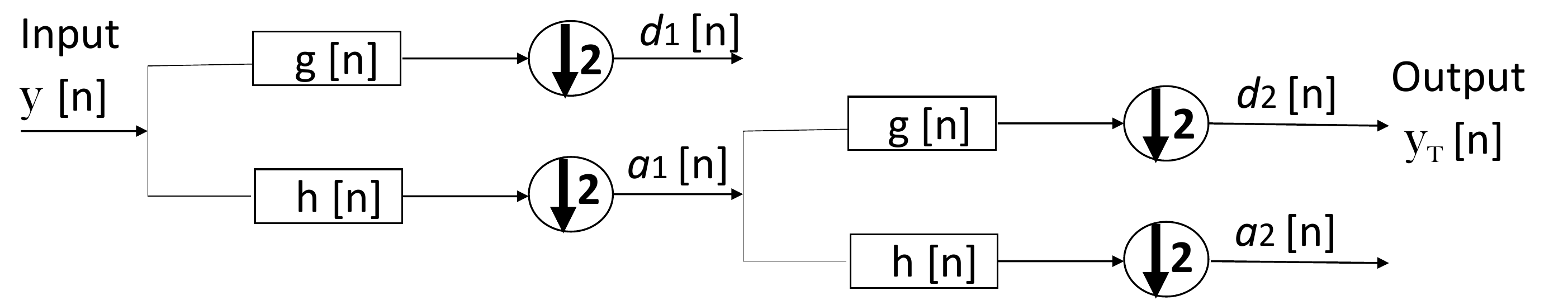}
    \caption{The two-level discrete Haar wavelet transform for pre-processing of the captured raw data.}
    \label{fig:Wavelet}
    \vspace{-3mm}
\end{figure}

In this work, multiresolution decomposition of the captured RF data are carried out using the two-level Haar transform as shown in Fig.~\ref{fig:Wavelet}. Using this transform, the raw input signal is decomposed into subbands, and important time-frequency information can be extracted at different resolution levels~\cite{bultan2000system}. In the first level, the input RF data are split into low- and high-frequency components by means of the half band low-pass ($h[n]$) and high-pass ($g[n]$) filters, respectively. This process is followed by a dyadic decimation, or downsampling, of the filter outputs to produce the approximate coefficients, $a_{1}[n]$, and detail coefficients, $d_{1}[n]$. In the second level, $a_{1}[n]$ coefficients are further decomposed in a similar manner, and the generated $d_{2}[n]$ coefficients are taken as the final output ($y_{\textrm{T}}[n]$). Then, $y_{\textrm{T}}[n]$ is input to the multistage detection system. Moving from left to right in Fig.~\ref{fig:Wavelet}, we get coarser representation of the captured RF data. The output RF data will have fewer samples due to the successive downsampling of the input RF data. This reduces the computational complexity of the overall process. Multiresolution analysis is also useful in detecting weak signals in the presence of background noise and removing the bias in the signals, leading to a higher detection accuracy, which is required in applications like UAV threat detection.\looseness=-1
\begin{figure}[t!]
\center{
 \begin{subfigure}[]{\includegraphics[trim=0.40cm 0cm 0.63cm 0.3cm, clip,width=0.47\linewidth]{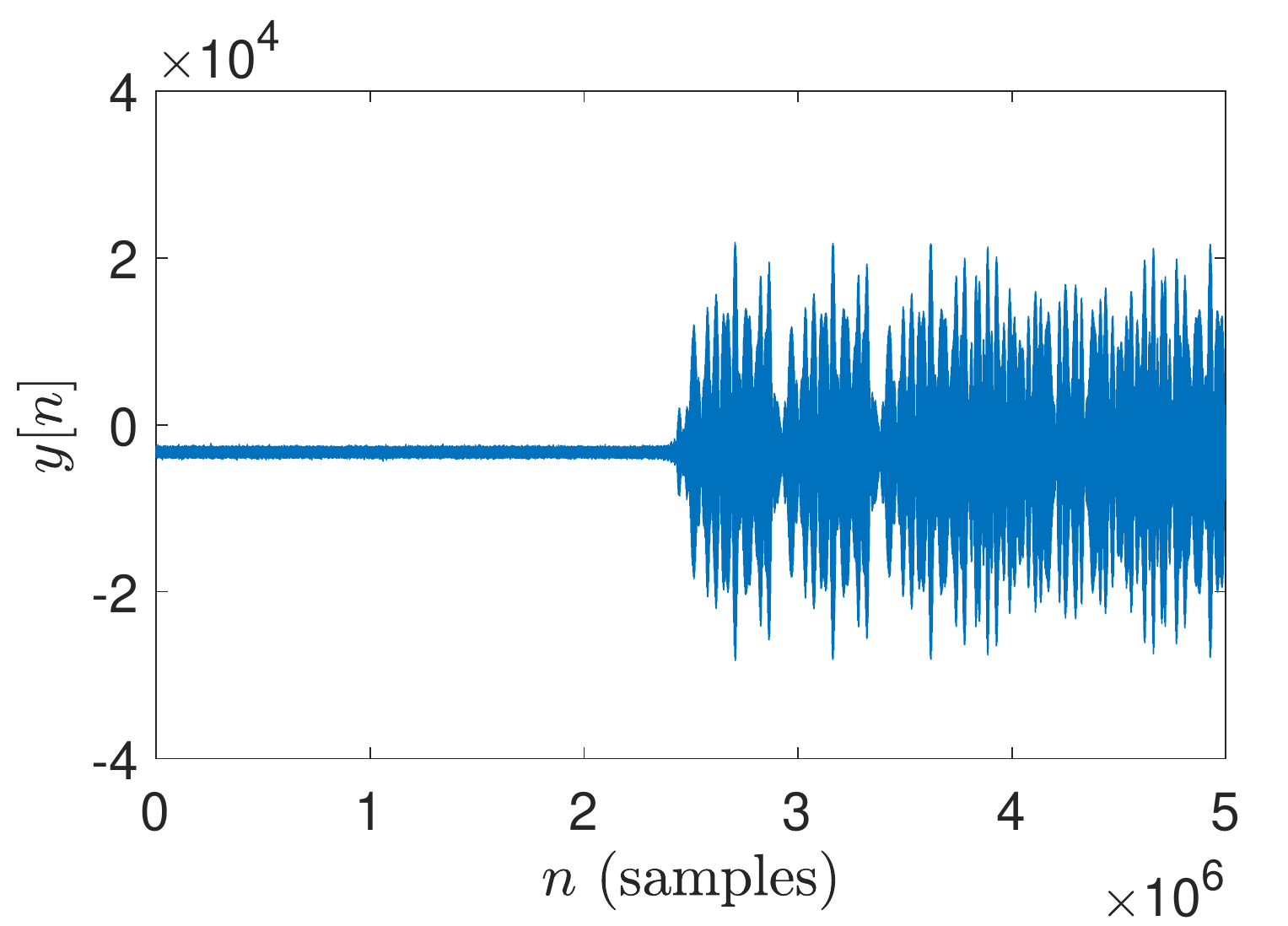}\label{fig:phantom}}
\end{subfigure}
 \begin{subfigure}[]{\includegraphics[trim=0.2cm 0.12cm 0.7cm 0cm, clip,width=0.47\linewidth]{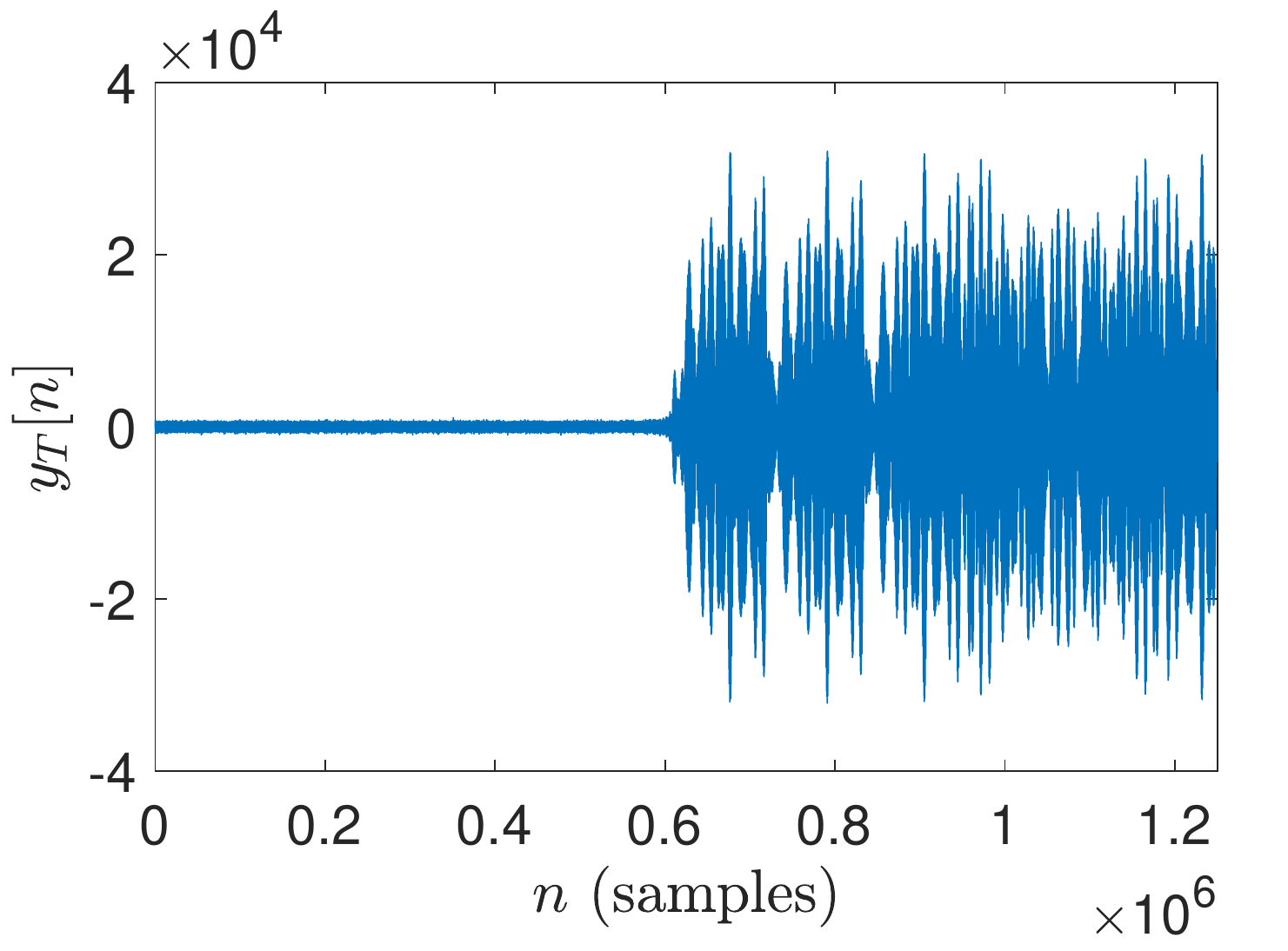}\label{fig:phantom_haart}}
\end{subfigure}
 \caption{(a) The sampled raw data $y[n]$ captured from the remote controller of a DJI Phantom 4 Pro UAV using an oscilloscope with a sampling rate of 20~Gsa/s, and (b) the transformed data $y_{\textrm{T}} [n]$ obtained at the output of the two-level Haar wavelet filter. Due to successive downsampling,  $y_{\textrm{T}} [n]$ has about $3.8\times 10^6$ fewer data samples than $y[n]$.}
 \label{fig:Haart_Example}}
 \vspace{-3mm}
 \end{figure}

Fig.~\ref{fig:Haart_Example} shows the effect of the Haar wavelet decomposition on a sample signal captured from the controller of the DJI Phantom 4 Pro UAV. It is clear from the figure that the wavelet transform removes the bias in the signal alignment and reduces the data size. It will be shown in Section~\ref{four_1}, the transformation also preserves the characteristics of the original waveform. After the pre-processing step, the data is transferred to the first stage of the detection system, where we decide if the captured data is an RF signal or noise.

\subsection{Na\"ive Bayes Decision Mechanism for RF Signal Detection}
In this stage, we first model the pre-processed RF data, $y_{\textrm{T}} [n]$, using two-state Markov models for ``RF signal" and ``noise" classes. This allows us to compute the likelihood that the captured data come from either the signal or noise class. According to the Bayesian decision theory, the optimum detector is the one that maximizes the posterior probability. Mathematically, let ${C} \in \{0,1\}$ be an index denoting the class of the pre-processed RF data $y_{\textrm{T}} [n]$, where ${C} = 1$ when the captured raw signal $y[n]$ is an RF signal, and ${C} = 0$ otherwise. Let ${S}_{y_{\textrm{T}}}=[{S}_{y_{\textrm{T}}}(1), {S}_{y_{\textrm{T}}}(2),\dots, {S}_{y_{\textrm{T}}}(N)]^\top$ be the state vector representation of the given test data ${y_{\textrm{T}}} [n]$ containing $N$ samples, with ${S}_{y_{\textrm{T}}}(i)\in\{S_1,S_2\}$, $i=1,2,..,N$, and $S_1$ and $S_2$ being the two states in the Markov models.
Then, the posterior probability of the RF signal class given ${S}_{y_{\textrm{T}}}$ is \looseness=-1
\begin{equation}
    P({C}=1|{S}_{y_{\textrm{T}}}) = \frac{P({S}_{y_{\textrm{T}}} | {C}=1) P({C}=1)}{P({S}_{y_{\textrm{T}}})},
    \label{Bayes_equation}
\end{equation}
where $P({S}_{y_{\textrm{T}}} | {C}=1)$ is the likelihood function conditioned on ${C}=1$, $P({C} = 1)$ is the prior probability of the RF signal class, and $P({S}_{y_{\textrm{T}}})$ is the evidence. A similar expression holds for the posterior probability $P({C}=0|{S}_{y_{\textrm{T}}})$. In practice, since the evidence is not a function of $C$, it can be ignored. Therefore, we are only interested in maximizing the numerator in~(\ref{Bayes_equation}). That is,
\begin{align}
   \widehat{C}= \arg\max_C {P({S}_{y_{\textrm{T}}}|{C})P({C})}.
\end{align}



We decide that the captured signal belongs to an RF signal (i.e., ${C} = 1$), if
\begin{equation}
    P({S}_{y_{\textrm{T}}}|{C}=1)P({C}=1) \geq P({S}_{y_{\textrm{T}}}|{C}=0)P({C}=0).
    \label{Naivedecision_rule}
\end{equation}
If we assume the prior probabilities of the signal and noise class are equal, then the decision rule in~(\ref{Naivedecision_rule}) reduces to
\begin{equation}
    P({S}_{y_{\textrm{T}}}|{C}=1) \geq P({S}_{y_{\textrm{T}}}|{C}=0).
\end{equation}
Therefore, for a given test data, we need to compute and compare the likelihood probabilities $P({S}_{y_{\textrm{T}}}|{C}=\{0,1\})$. First, in order to compute the likelihood probability for the RF signal and noise classes, we use large amount of training data captured from multiple UAV controllers, Wi-Fi routers, mobile Bluetooth emitters, and background noise. This training data set is stored in a database as shown in Fig.~\ref{Fig:flowchart}. Since the captured RF data (after sampling) is a discrete time-varying waveform, we can model it as a stochastic sequence of states/events. The likelihood probability of such a state sequence can be computed based on the transitions between the states of the generated Markov models.



\begin{figure}[t!]
\center{
 \begin{subfigure}[]{\includegraphics[scale=0.33]{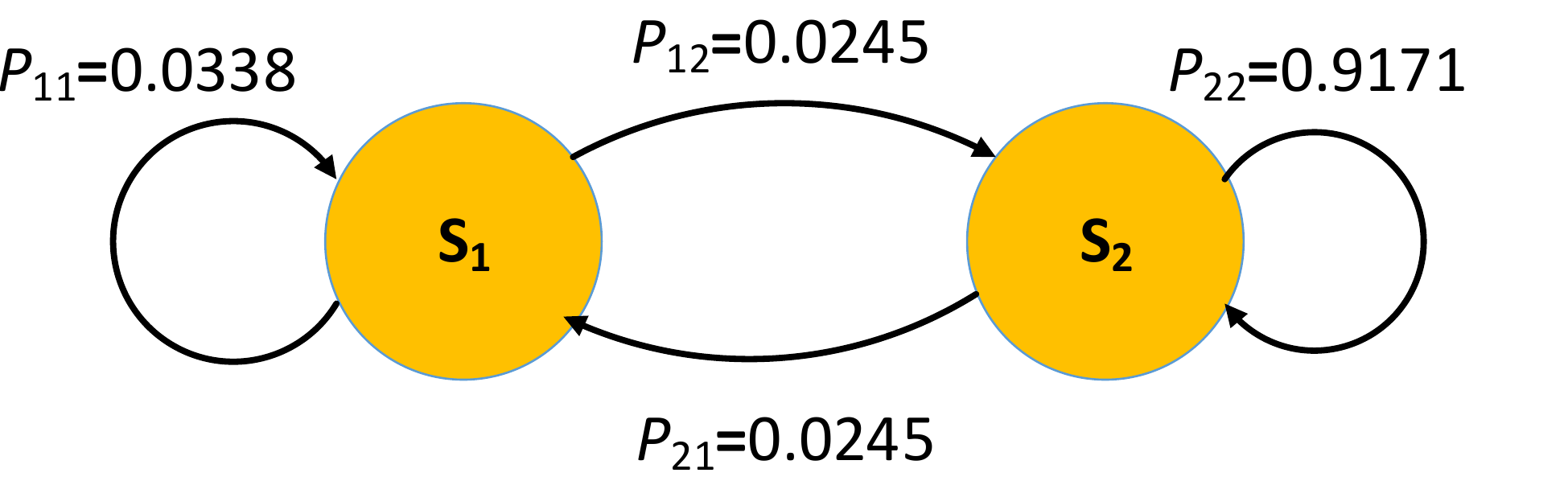}\label{fig:Markov_signal_class}}
\end{subfigure}
 \begin{subfigure}[]{\includegraphics[trim=0.1cm 0cm 0cm 0.8cm, clip,scale=0.35]{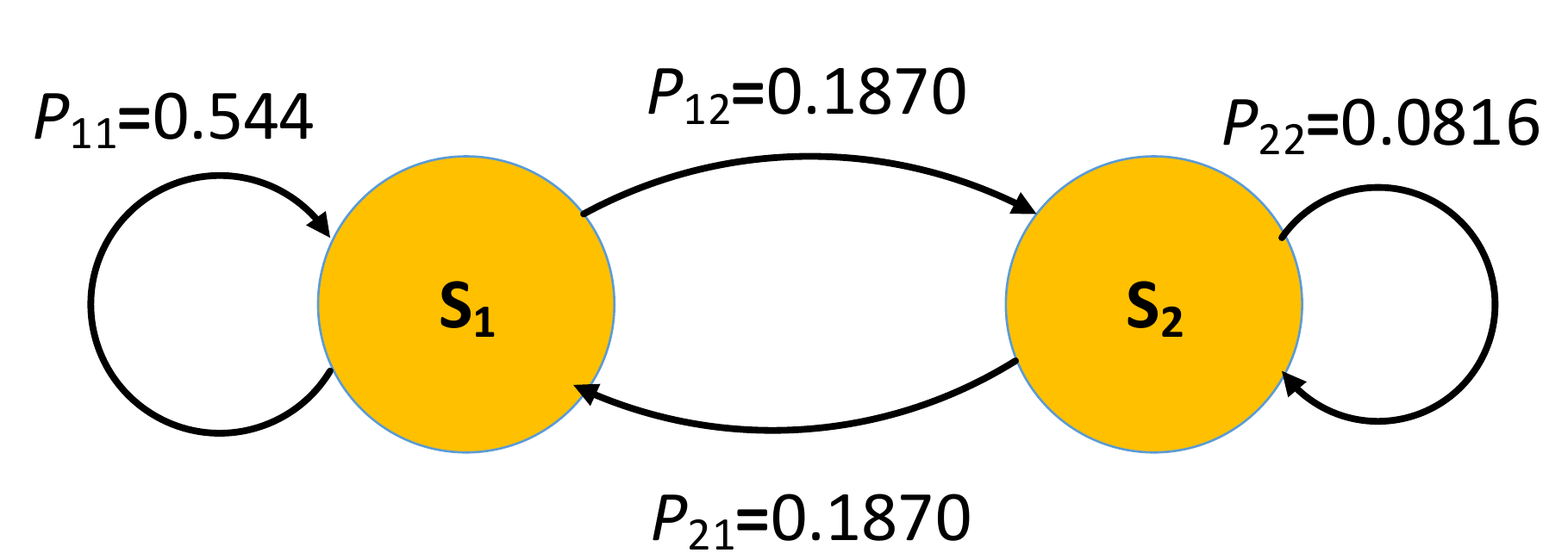}\label{fig:Markov_noise_class}}
\end{subfigure}
 \caption{Two-state Markov model and associated state transition probabilities using $\delta=3.5\sigma$ for (a) the RF signal class, and (b) the noise class.}
 \label{fig:transition}}
 \vspace{-3mm}
 \end{figure}




A two-state Markov model for a given signal $y_{\textrm{T}}[n]$ can be generated by mapping each sample in the signal to one of the two states ($S_1$ and $S_2$). The samples whose absolute amplitudes are less than or equal to a predetermined threshold $\delta$ are considered as in state $S_1$, while the samples with absolute amplitude greater than $\delta$ are considered as in state $S_2$. Mathematically, the state transformation is performed as follows:
\begin{equation}\label{equ_state_transition}
{S}_{y_{\textrm{T}}}(n)=
\begin{cases}
S_1, & \left|{y_{\textrm{T}}[n]}\right|\leq \delta \\
S_2, & \left|{y_{\textrm{T}}[n]}\right|>\delta
\end{cases}.
\end{equation}
Based on the above rule, it is straightforward to transform $y_{\textrm{T}}[n]$ into the state vector, ${S}_{y_{\textrm{T}}}$. Once ${S}_{y_{\textrm{T}}}$ is obtained, the probability of a transition between any two states is calculated. Note that the state vector is generated based on the amplitude of the signal samples in the wavelet domain. The choice of $\delta$ in~(\ref{equ_state_transition}) depends on the operating SNR of the system and will be discussed in Section~\ref{results_detection}. The transition count matrix, $\boldsymbol{T}_N$, and the transition probability matrix, $\boldsymbol{T}_P$, are defined as follows:
\begin{equation}
\boldsymbol{T}_N =
\begin{bmatrix}
    N_{11}       & N_{12}   \\
    N_{21}       & N_{22}   \\
\end{bmatrix},
\boldsymbol{T}_P =
\begin{bmatrix}
    p_{11}       & p_{12} \\
    p_{21}       & p_{22}   \\
\end{bmatrix} = \frac{\boldsymbol{T}_N}{\sum_{i,j}^{} N_{ij}},
\end{equation}
respectively, where $N_{ij}$ is the number of transitions from state $S_i$ to $S_j$ among all samples of $y_\textrm{T}[n]$, and $p_{ij} = P(S_i \rightarrow S_j)$ is the probability of a transition from state $S_i$ to $S_j$. The matrix $\boldsymbol{T}_P$ is obtained by normalizing the $\boldsymbol{T}_N$ matrix with the total number of samples in the signal. It is expected that the transition probabilities generated for the signal class (UAV, Wi-Fi, and Bluetooth) and the noise class will be significantly different at modest SNR levels. Also, the choice of $\delta$ in (\ref{equ_state_transition}) dictates the transition probabilities for both the signal and noise class. In Section~\ref{results_detection}, the threshold $\delta$ is expressed in terms of the estimated standard deviation ($\sigma$) of the preprocessed noise data captured from the environment. Moreover, during the experiments, data is captured within a short-time window (0.25~ms), thus we assume the environmental noise is stationary during this interval.

Fig.~\ref{fig:transition} shows the two-state Markov models for the RF signal and noise classes obtained from the training data using $\delta=3.5\sigma$. From Fig.~\ref{fig:Markov_signal_class}, we see that for the signal class, $p_{22}$ is significantly higher than $p_{11}$, $p_{12}$, and $p_{21}$. On the other hand, from Fig.~\ref{fig:Markov_noise_class}, we see that for the noise class, $p_{11}$ is significantly higher than the other transition probabilities. Based on these observations, the differences between the state transition probabilities of each class can be utilized to determine the class of a captured test signal.



\begin{figure*}{}
\center{
 \begin{subfigure}[]{\includegraphics[scale=0.4]{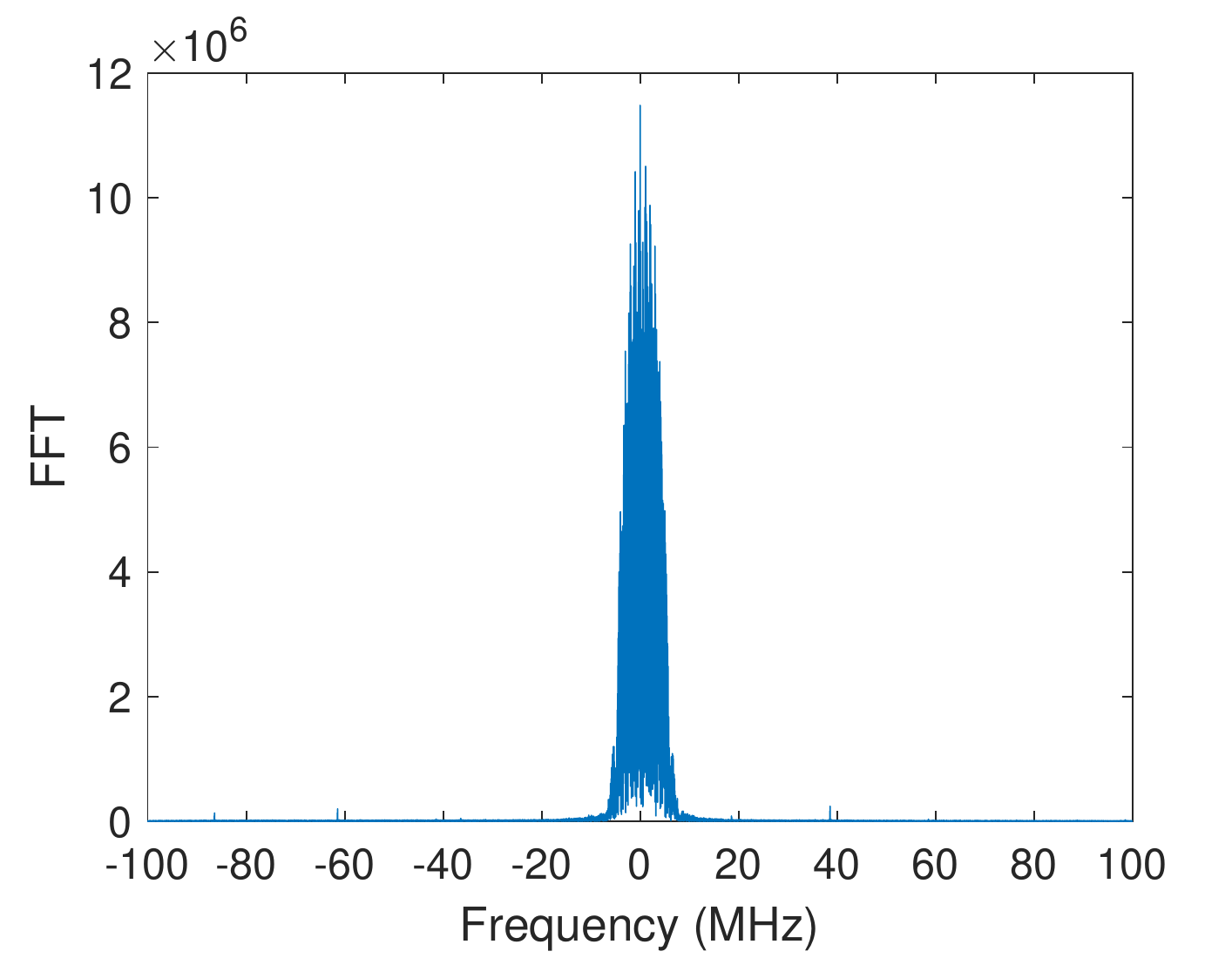}\label{Wifi_BW}}
\end{subfigure}
 \begin{subfigure}[]{\includegraphics[scale=0.4]{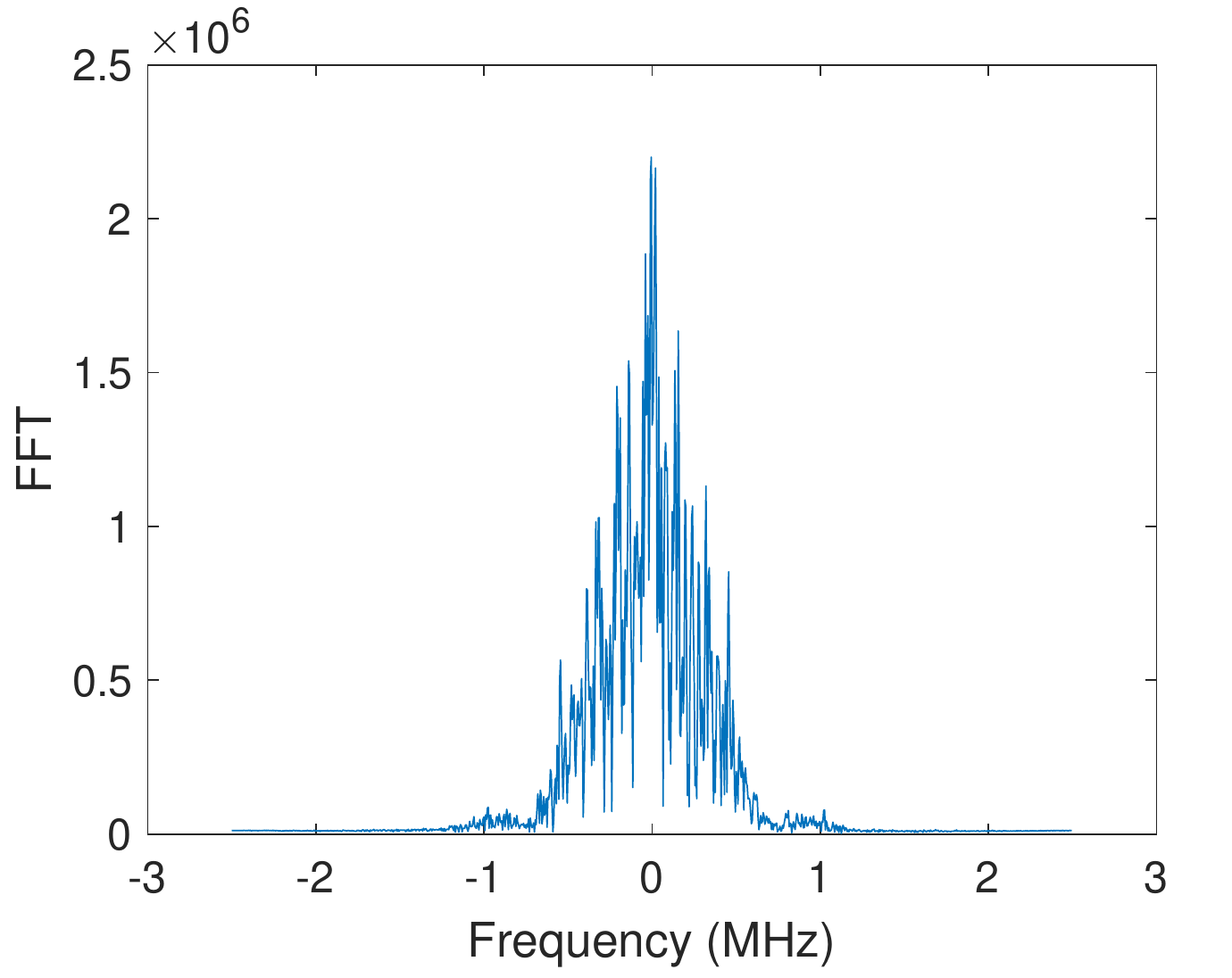}\label{Motorola_BW}}
\end{subfigure}
\begin{subfigure}[]{\includegraphics[scale=0.4]{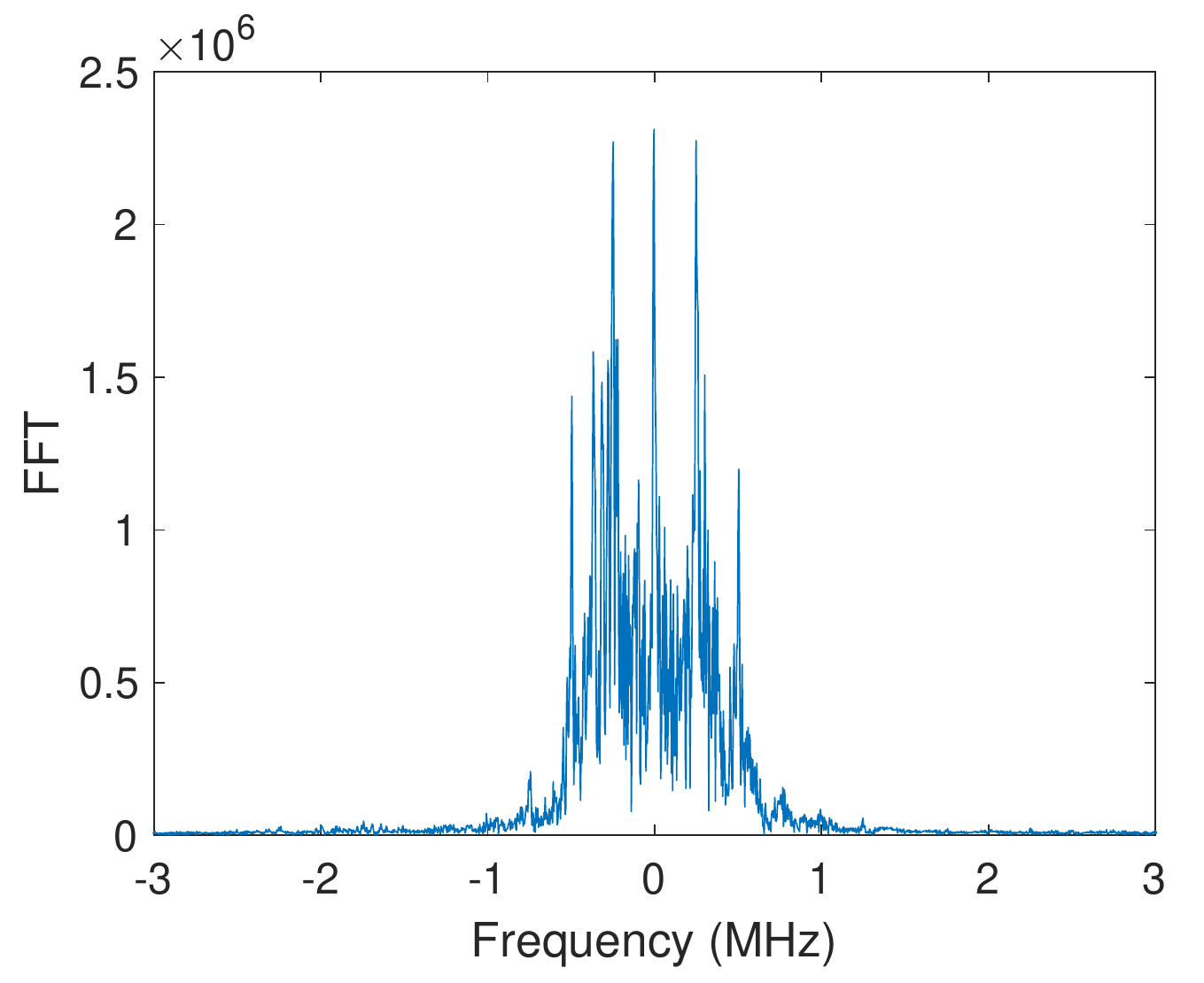}\label{FSK_controller_BW}}
\end{subfigure}

 \caption{{Bandwidth analysis of (a) Wi-Fi signal, (b) Bluetooth signal from Motorola e5 cruise, and (c) Spektrum DX5e UAV controller signal.}}
 \label{fig:Bandwidth_analysis}}
 \vspace{-3mm}
 \end{figure*}


Consequently, the likelihood of the test signal being an RF signal can be calculated as follows:
\begin{eqnarray}
\begin{aligned}
      P({S}_{y_{\textrm{T}}}|{C}=1) &= \prod_{n=1}^{N-1} p({S}_{y_{\textrm{T}}}(n) \rightarrow {S}_{y_{\textrm{T}}}(n+1) | {C}=1)\\
           &= \prod_{i,j=\{1,2\}} {\boldsymbol{T}_{P_{{C}=1}}^{\boldsymbol{T}_N(i,j)}(i,j)}\\
     &= \prod_{i,j=\{1,2\}} p_{ij;{{C}=1}}^{N_{ij}}.
\end{aligned}
\end{eqnarray}
The product of the conditional transition probabilities in the above equation gives the likelihood of obtaining the state vector ${S}_{y_{\textrm{T}}}$ given the hypothesis $C=1$ is true. The log-likelihood of the above expression is:
\begin{eqnarray}
\begin{aligned}
     \log\left( P({S}_{y_{\textrm{T}}}|{C}=1)\right) &= \sum_{i,j=\{1,2\}} N_{ij} \log(p_{ij;{{C}=1}}).
\end{aligned}
\end{eqnarray}
Similarly, the log-likelihood of the signal coming from a noise class is calculated by
\begin{eqnarray}
\begin{aligned}
     \log\left( P({S}_{y_{\textrm{T}}}|{C}=0)\right) &= \sum_{i,j=\{1,2\}} N_{ij} \log(p_{ij;{{C}=0}}).
\end{aligned}
\end{eqnarray}
The decision will be favored to $ {C}=1$, if $\log(P({S}_{y_{\textrm{T}}}| {C}=1)) \geq \log(P({S}_{y_{\textrm{T}}}|{C}=0))$; otherwise, ${C}=0$. We discuss the detection results in Section~\ref{results_detection}. If the captured test signal belongs to the RF signal class, then the second stage detector is invoked to identify UAV controller-type signals. Otherwise, the system continues sensing the environment for the presence of signals as shown in Fig.~\ref{Fig:flowchart}.\looseness=-1

\section{Detection of Wi-Fi And Bluetooth Interference}\label{second_stage_detector}
In recent times, there has been interest in detecting Wi-Fi and Bluetooth signals~\cite{rayanchu2011airshark}. However, in the context of RF-based UAV detection in urban environments, where the Wi-Fi and Bluetooth signals are considered as interference, there has been minimal research efforts. Fortunately, these interference signals are well standardized and can be identified by using the knowledge of their specifications. Table~\ref{Table1} provides a brief summary of the specifications for Wi-Fi and Bluetooth transmissions. It is obvious that the signal bandwidth and the modulation type are two important features for identifying the Wi-Fi and Bluetooth signals. The second stage detector exploits these features for detecting these interference sources.

The first step in deciding if the detected signal is a wireless interference or not is to perform bandwidth analysis. This is because Wi-Fi signals can be easily identified by their bandwidth. According to Table~\ref{Table1}, Bluetooth 2.0 signals have a bandwidth of 1 or 2~MHz, Wi-Fi signals have a bandwidth of 20~MHz (or more) while all the UAV controller signals in our database have bandwidth less than 10~MHz. Therefore, if the detected RF signal has a bandwidth equal or greater than 20~MHz, it is classified as a Wi-Fi signal. Bandwidth analysis is performed by taking the Fourier transform of the resampled signal. Fig.~\ref{fig:Bandwidth_analysis} shows the result of the bandwidth analysis of a typical Wi-Fi, a Bluetooth (from Motorola e5 cruise), and a UAV (Spektrum DX5e) controller signal.
 \begin{table}[t]
\centering
\caption{Specifications of the Wi-Fi and Bluetooth Standards.}
\label{Table1}
\begin{tabular}{|p{3.5cm}| m{2cm}|m{2cm}|m{2cm}|}
\hline
Standard & Bluetooth (IEEE 802.15.1 WPAN) & Wi-Fi (IEEE 802.11 WLAN)\\
\hline
Center frequency (GHz)& 2.4 & 2.4/ 5 \\
Bandwidth (MHz) & 1 & 20/ 40/ 80/ 160 \\
PHY modulation\cite{Bluetooth_resource,Wi-Fi_resource}  & GFSK/FSK/DPSK &  DSSS/ OFDM\\
 Range (m)& variable & >50\\
Data rate (Mbps) & 2 & variable \\
\hline
\end{tabular}
\vspace{-3mm}
\end{table}

 \begin{figure*}{}
\center{
 \begin{subfigure}[]{\includegraphics[scale=0.37]{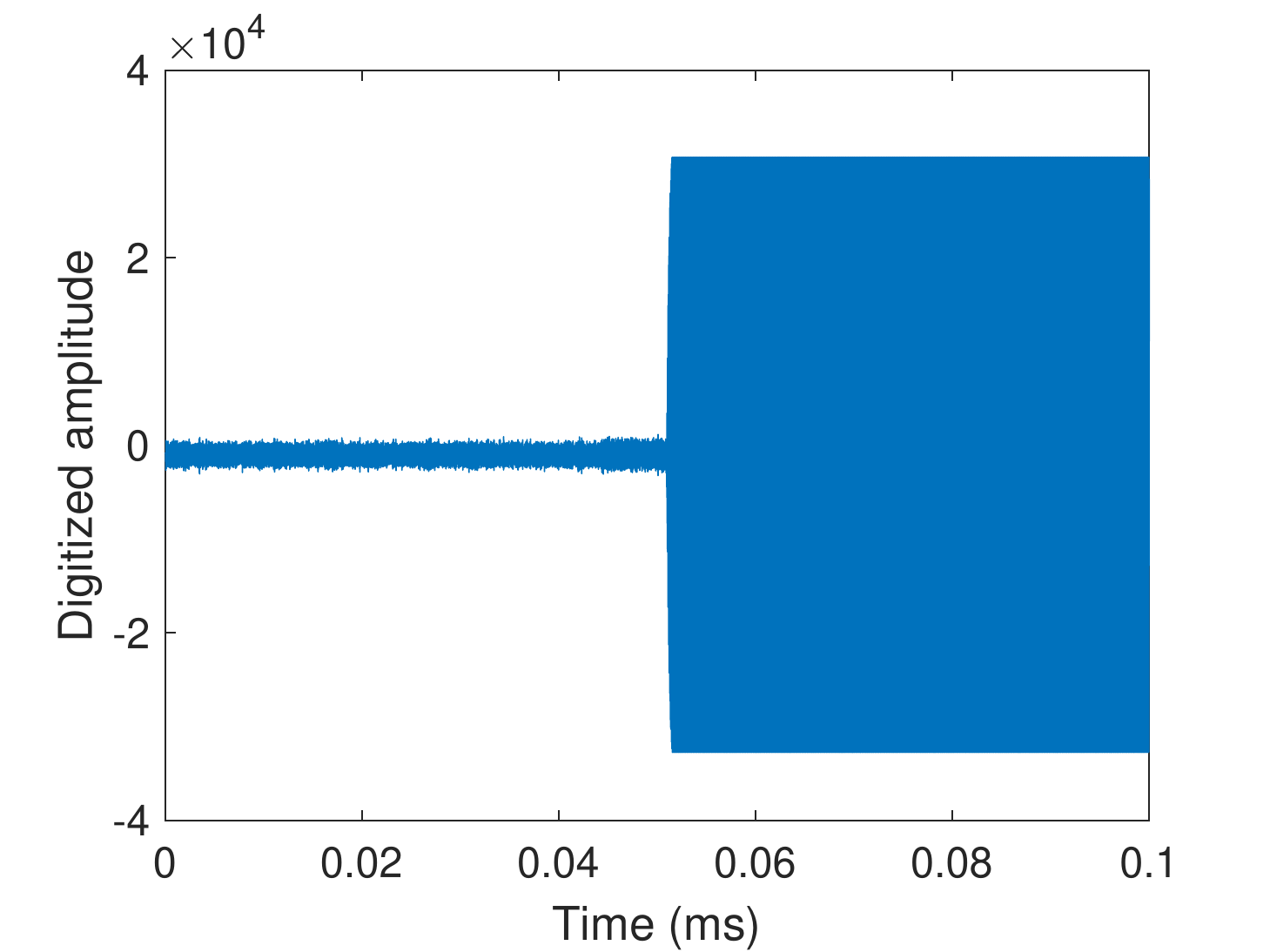}\label{Motorola_Raw}}
\end{subfigure}
 \begin{subfigure}[]{\includegraphics[scale=0.37]{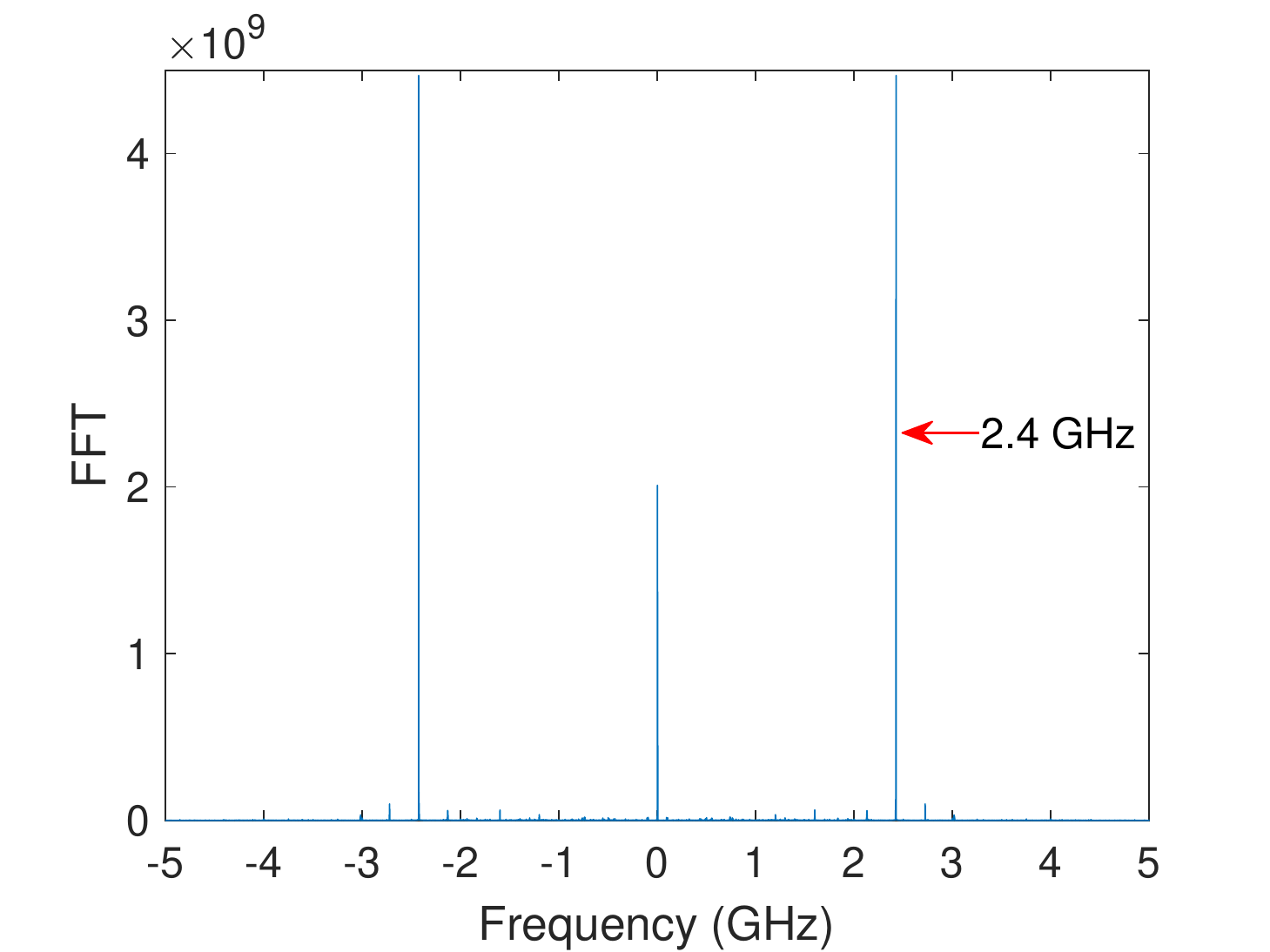}\label{Motorola_FFT_Raw}}
\end{subfigure}
\begin{subfigure}[]{\includegraphics[scale=0.37]{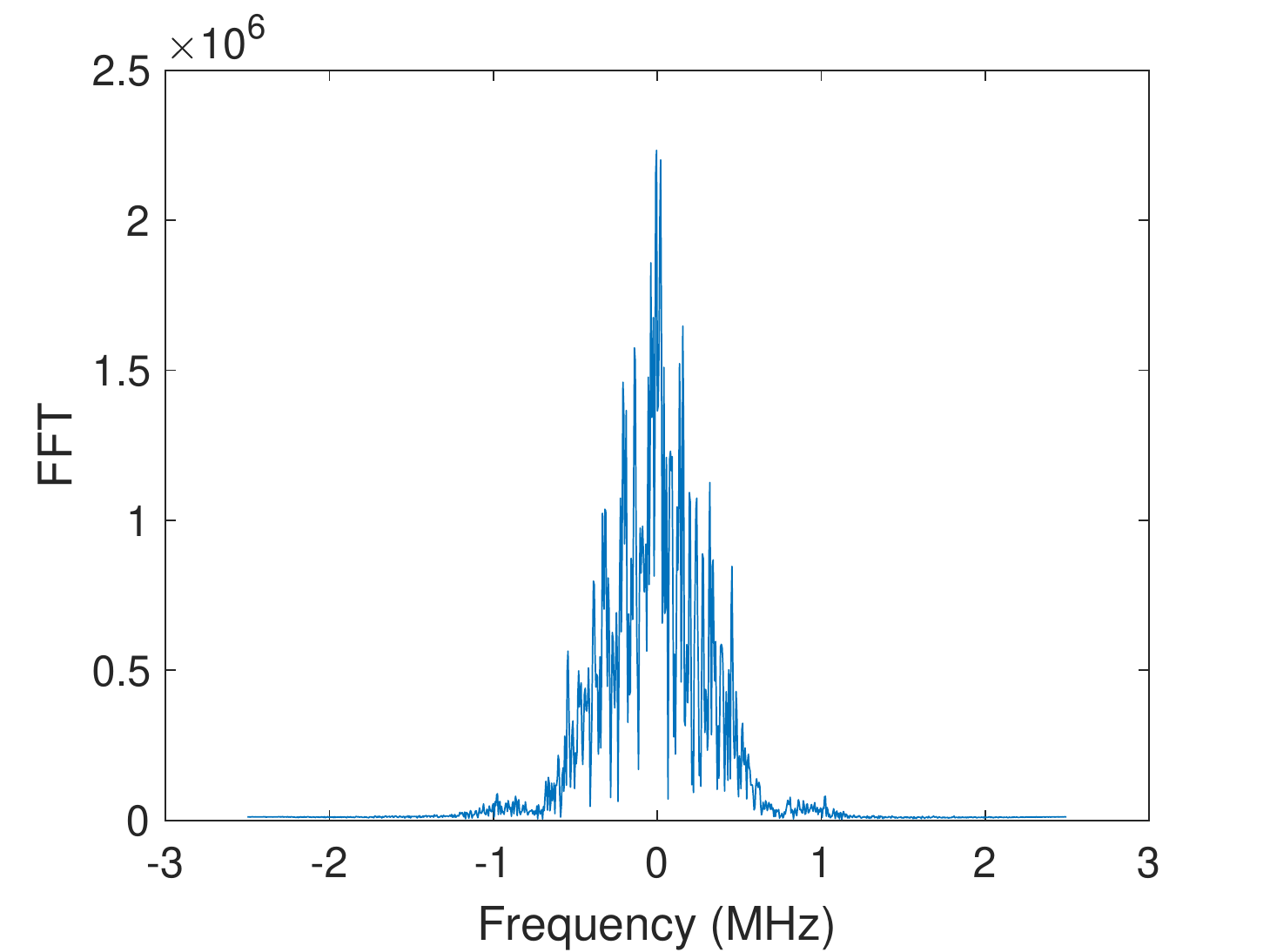}\label{Motorola_Resampled}}
\end{subfigure}
\hfill
\begin{subfigure}[]{\includegraphics[scale=0.37]{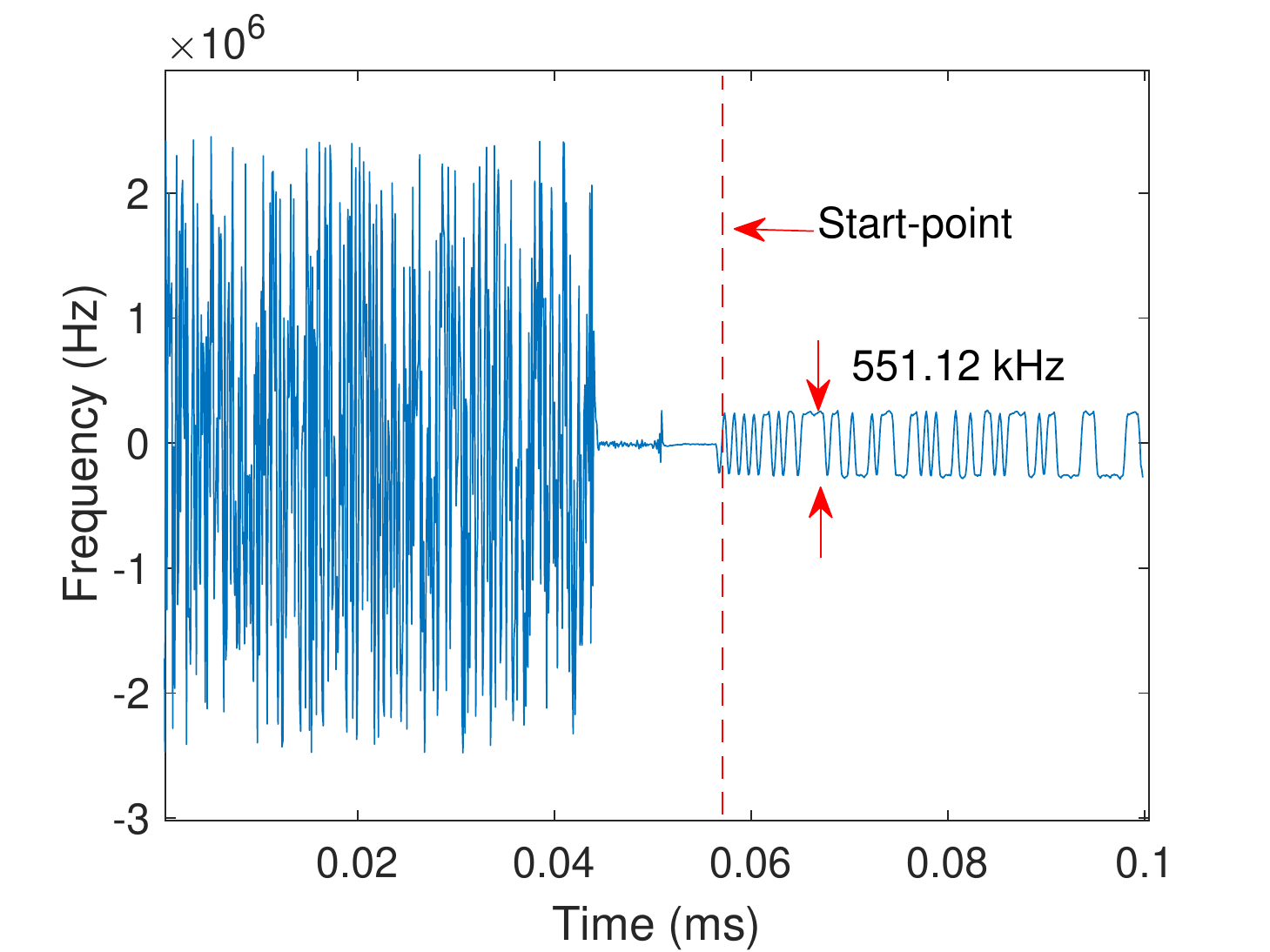}\label{Motorola e5 cruise mobile device}}
\end{subfigure}
 \begin{subfigure}[]{\includegraphics[scale=0.37]{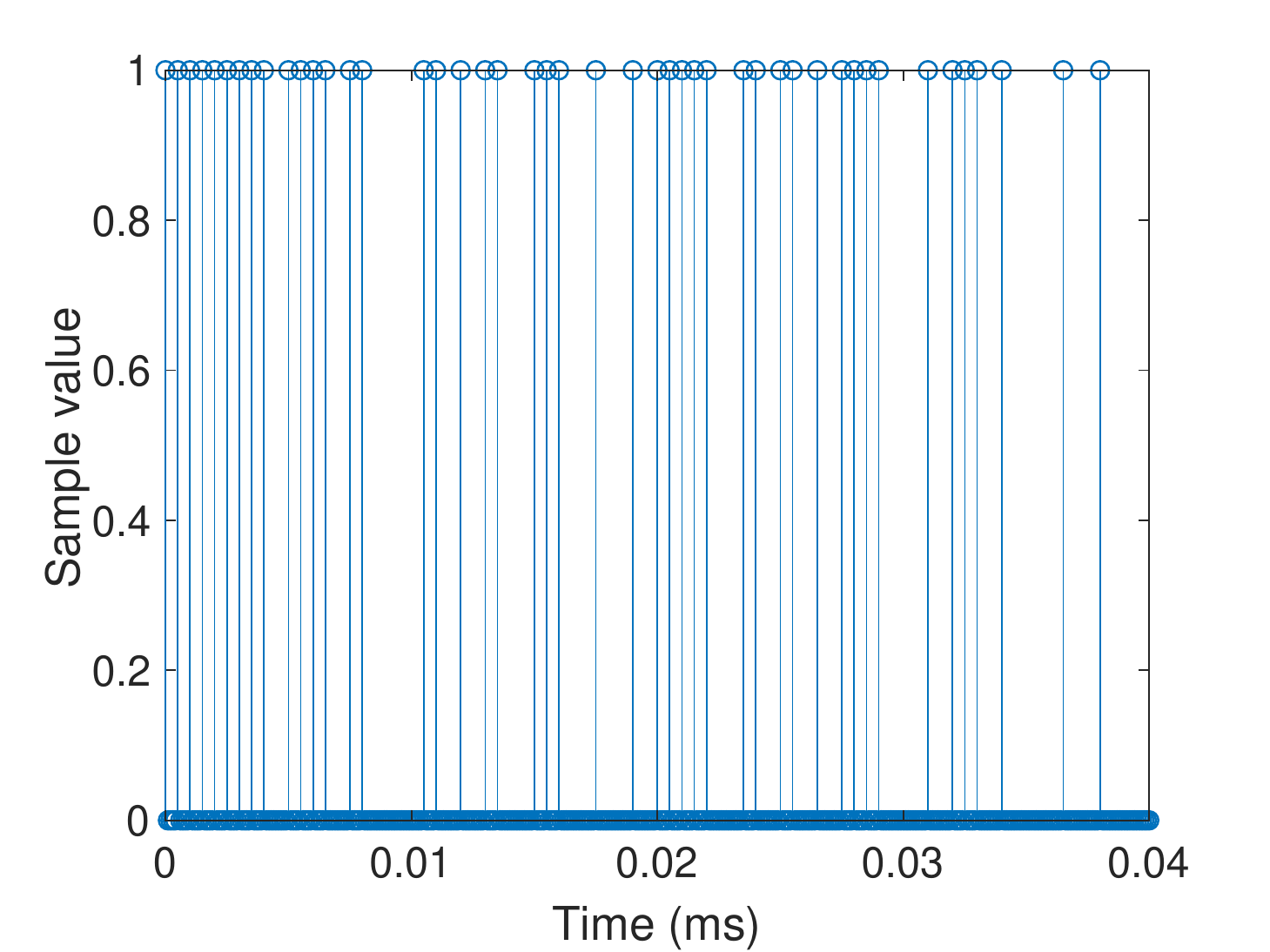}\label{Motorola_binary}}
\end{subfigure}
\begin{subfigure}[]{\includegraphics[scale=0.37]{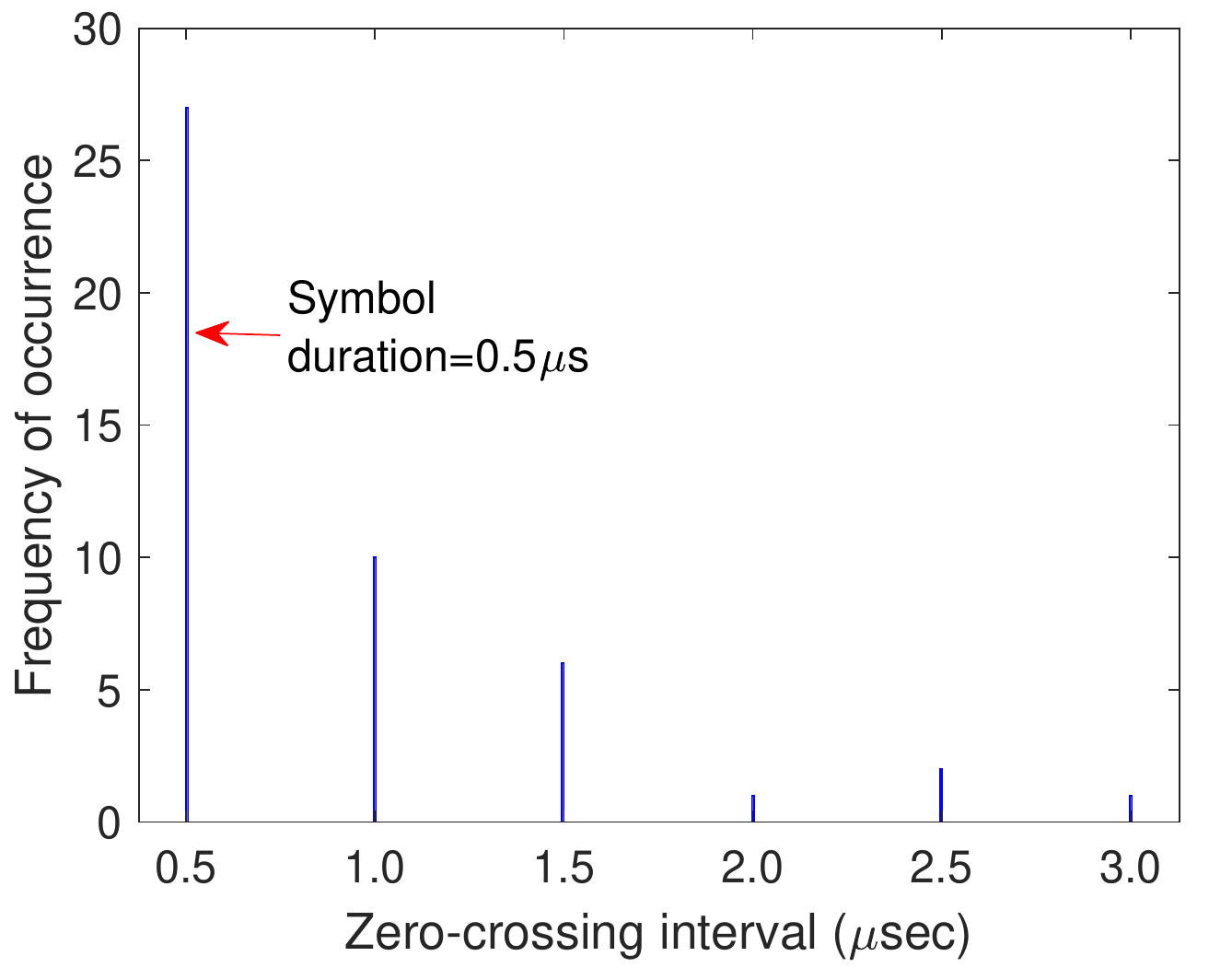}\label{Motorola_histogram}}
\end{subfigure}
 \caption{{Extraction of the modulation features of a Bluetooth interference signal from Motorola e5 cruise mobile device using zero-crossing demodulation technique: (a) Raw signal, (b) FFT of the raw signal, (c) FFT of the shifted and resampled signal (by 1/2000), (d) the demodulated signal showing a peak-to-peak frequency of 551.12~kHz, (e) binary signal, and (f) histogram of the time-interval between consecutive zero-crossings in the modulated signal.}}
 \label{fig:Bluetooth_demodulation}}
 \vspace{-3mm}
 \end{figure*}



If the  detected signal has a bandwidth less than 20 MHz, it is assumed to be transmitted either from a Bluetooth interference source or a valid UAV controller. Since most mobile Bluetooth devices employ Gaussian frequency shift keying GFSK/FSK modulation, it is reasonable to detect and discriminate these devices by means of modulation features. In this study, two GFSK/FSK modulation features, namely, frequency deviation and symbol duration, will be used to discriminate Bluetooth signals. Frequency deviation is a measure of the maximum difference between the peak frequency in the GFSK/FSK signal and the center frequency. On the other hand, symbol duration is the minimum time interval in the observed Bluetooth waveform or pulse. Therefore, using a GFSK/FSK demodulator, these features can be extracted and used as the basis for Bluetooth signal detection.

We consider a zero-crossing GFSK/FSK demodulator. It is known that the Bluetooth GFSK/FSK signal is transmitted in burst consisting of $M$ data bits ${d_{m}} \in \{-1,+1\}$, each bit having a period $T_{b}$ and average energy per bit $E_{b}$~\cite{scholand2003bluetooth}. A general model for such a signal is given as:
 \begin{equation}
   s(t)=\sqrt{\frac{2E_{b}}{T_{b}}}.\cos (2\pi f_{o}t+\varphi(t,\alpha)+\varphi_{o}) + n(t),
   \label{GFSK}
\end{equation}
where $\varphi(t,\alpha)$ is a phase modulating function, $\varphi_{o}$ is an arbitrary phase constant,  $f_{o}$ is the operational frequency, and $n(t)$ is the channel noise component. The zero-crossing demodulator considered herein for Bluetooth interference detection is able to detect the time instants at which the signal $s(t)$ is equal to zero and has a positive slope, i.e., the zero-crossings. When a Bluetooth device transmits at the basic rate using the standard GFSK/FSK modulation, one symbol represents one bit. Therefore, the time interval between consecutive zero-crossings is a measure of the symbol duration of the Bluetooth signal. \looseness=-1


Fig.~\ref{fig:Bluetooth_demodulation} shows the results of the zero-crossing demodulation of a Bluetooth signal from Motorola e5 cruise. The captured Bluetooth signal and its fast Fourier transform (FFT) are shown in Fig.~\ref{Motorola_Raw} and Fig.~\ref{Motorola_FFT_Raw}, respectively. From Fig.~\ref{Motorola_FFT_Raw}, we see that the transmit frequency of the Bluetooth device is 2.4 GHz. Afterward, the signal is shifted and resampled by 1/2000. The FFT of the resampled signal is shown in Fig.~\ref{Motorola_Resampled}. This figure shows that the bandwidth of the Bluetooth signal is around 2~MHz, which is far less than 20~MHz. Next, the resampled signal is demodulated by taking the derivative of its phase angle, and the start point of the demodulated signal is estimated using the Higuchi algorithm~\cite{esteller2001comparison}. The Higuchi algorithm detects the start point of the signal by measuring the fractal dimensions of the signal. Once the start point is detected, the frequency deviation is estimated as one half the peak-to-peak frequency of the demodulated signal. Fig.~\ref{Motorola e5 cruise mobile device} shows a plot of the demodulated signal and the estimated start point which is obtained using the Higuchi algorithm. From the figure, the peak to peak frequency of the demodulated signal is estimated as 551.12~kHz, and therefore, the frequency deviation is 275.56~kHz.

In order to estimate the symbol duration, the demodulated signal is converted to a binary signal by using the mean as a threshold. Fig.~\ref{Motorola_binary} shows the binary signal, where binary one represents a positive frequency deviation, and a binary zero
represents a negative frequency deviation. Then, we compute the derivative of the binary signals to locate the zero-crossings. To ensure we accurately compute the symbol duration, we compute the histogram of the time intervals between consecutive zero-crossings. This is necessary because channel distortions will cause some deviation in these intervals. Fig.~\ref{Motorola_histogram} shows the histogram of the time intervals between consecutive zero-crossings for the Bluetooth signal. The estimated symbol duration is 0.5~$\mu$s.

\begin{figure}[]
\center
\includegraphics[trim=0.1cm 0cm 0.1cm 1.1cm, clip,width=0.42\textwidth]
{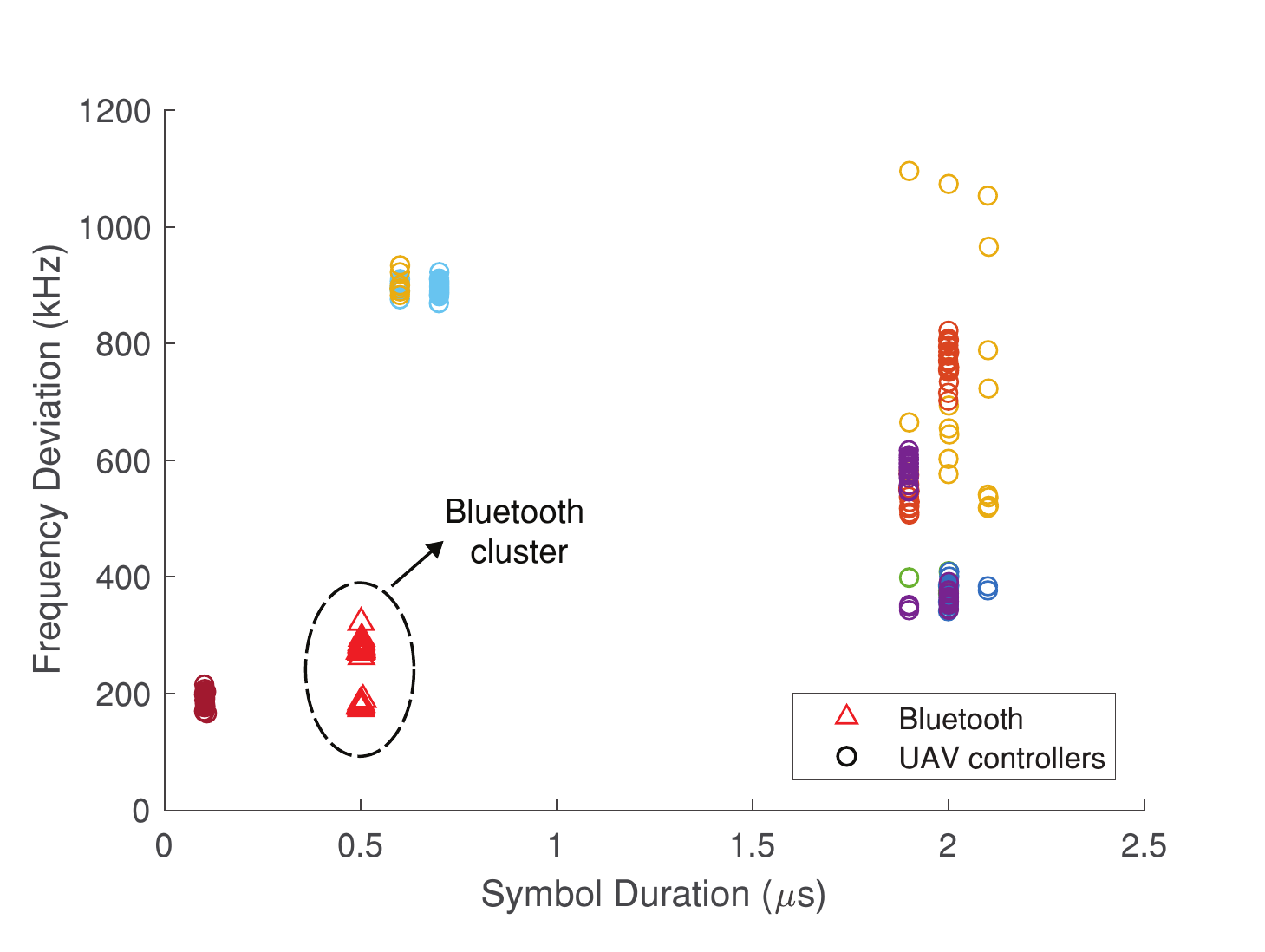}
\caption{Feature space showing the symbol duration and frequency deviation of the signals from several mobile Bluetooth devices and UAV controllers. Each UAV controller is represented by a circular marker of a unique color.}
\label{Bluetooth_VS_UAV}
\vspace{-3mm}
\end{figure}
To validate the joint discriminating ability of these modulation features, Bluetooth signals from six mobile phones and signals from nine UAV controllers are collected. The mobile phones are Iphone 7, Iphone XR, LG X charge, Motorola G Play, Motorola e5 cruise, and Samsung Galaxy Note 9. The UAV controllers considered are Jeti Duplex DC-16, Spektrum DX5e, Spektrum DX6e, Spektrum DX6i, Spektrum JR X9303, FlySky FS-T6, Graupner MC-32, HK-T6A, and Turnigy 9X. The UAV signals are frequency modulated as well. Therefore, all the collected signals are demodulated using the zero-crossing technique. Fig.~\ref{Bluetooth_VS_UAV} shows the feature space of the demodulated Bluetooth and UAV controller signals. The figure shows a clear clustering of the Bluetooth signals from different mobile phones. All the Bluetooth signals have a symbol duration of 0.5$\mu$s and a frequency deviation of less than 350~kHz. Therefore, the frequency deviation and symbol duration can be used as features in a simple maximum likelihood classifier for identifying Bluetooth interference signals. If the detected signal is not from a Bluetooth interference source, it is presumed to be an emission from a UAV controller and transferred to the UAV classification system.


\section{UAV Classification Using RF Fingerprints}\label{four_1}
The input to the ML classifiers are the RF-based features extracted from the energy-time-frequency domain representation of the UAV controller signals.
\begin{figure}{}
\center{
 \begin{subfigure}[]{\includegraphics[scale=0.5]{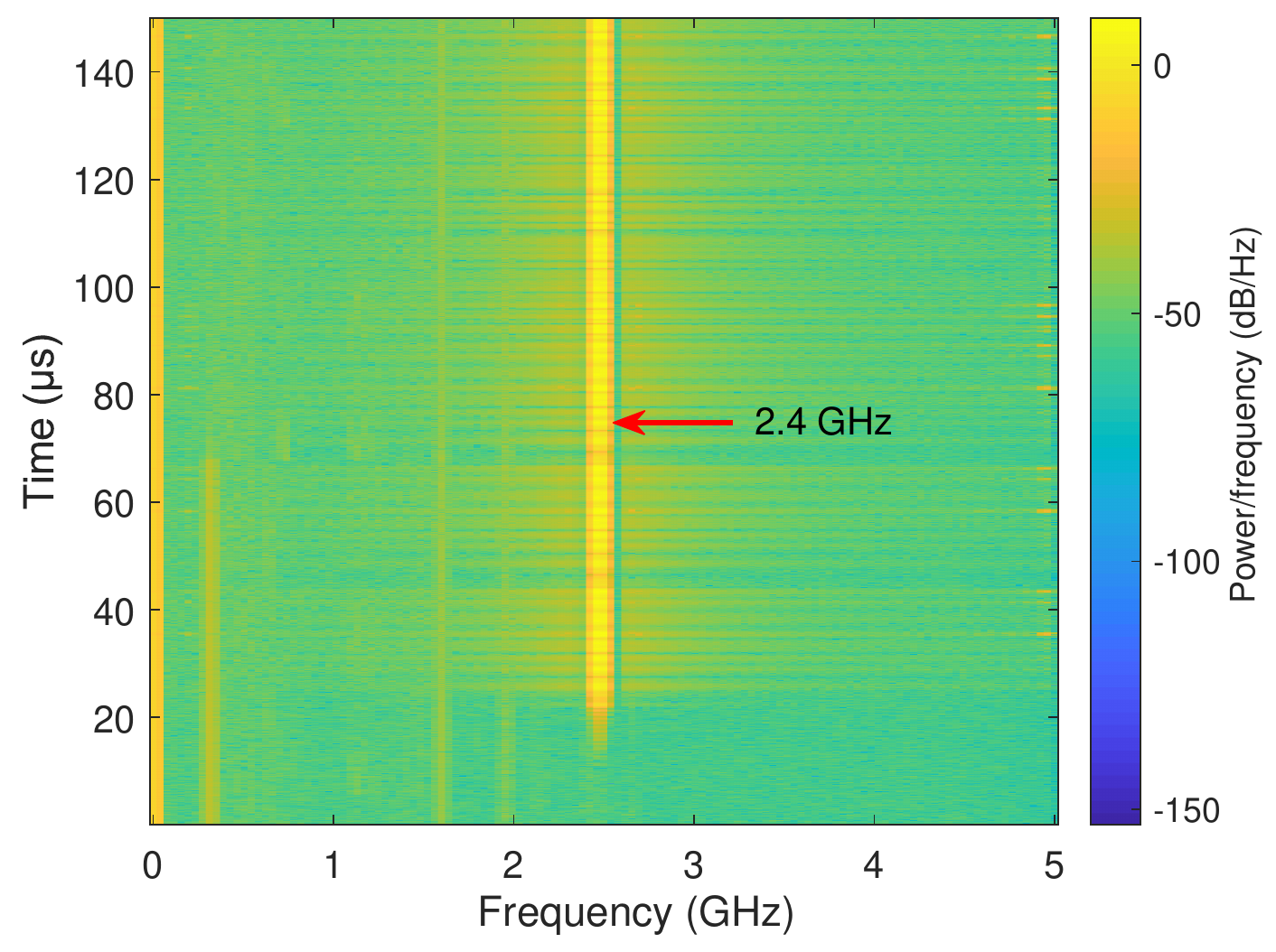}\label{fig:spectrogram}}
\end{subfigure}
\begin{subfigure}[]{\includegraphics[scale=0.5]{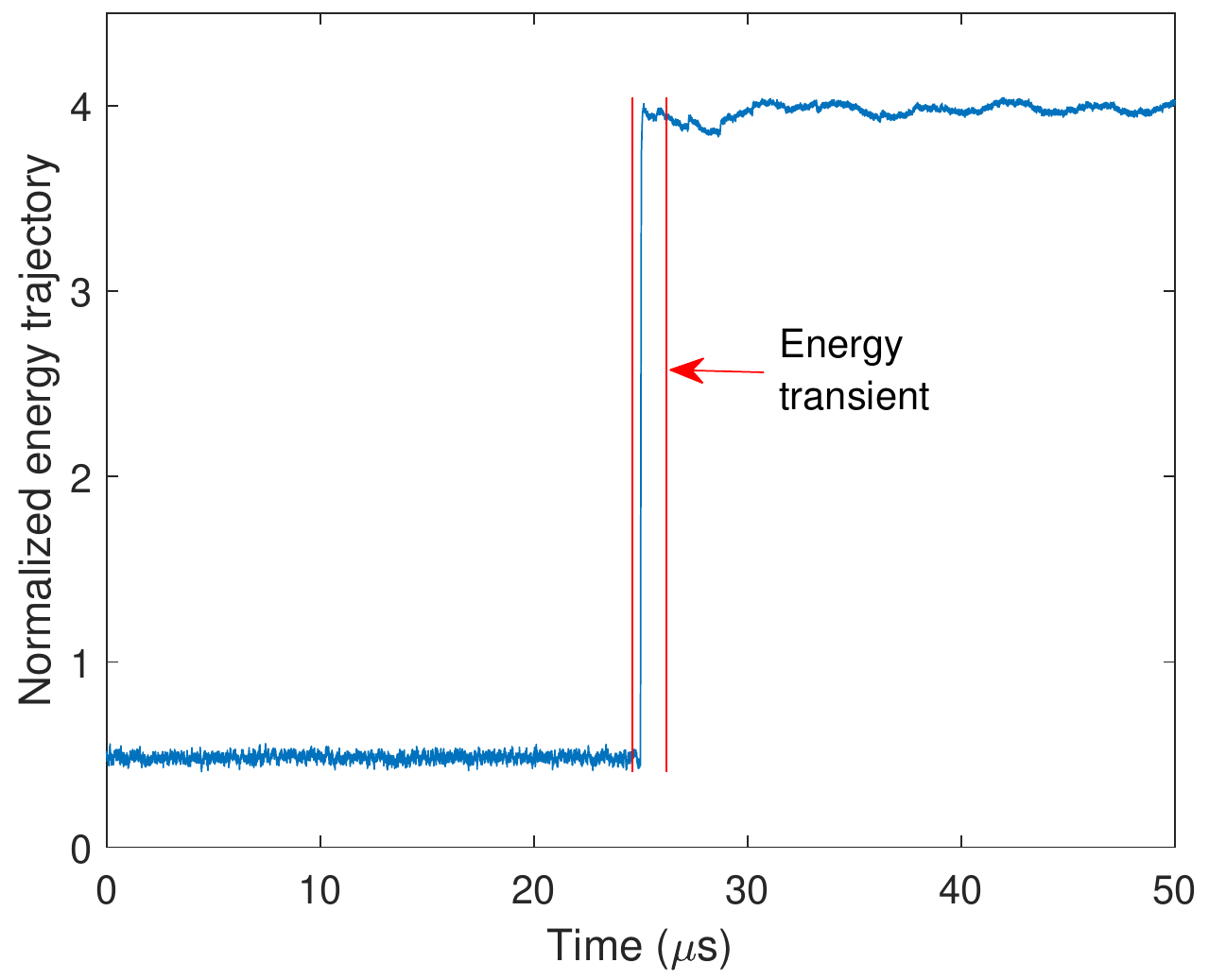}\label{fig:energy_trajectory}}
\end{subfigure}
 \caption{{(a) The spectrogram and, (b) the energy trajectory of the UAV controller signal shown in Fig.~\ref{fig:Haart_Example}.}}
\label{Fig:spectrogram_energy_trajectory}}
\vspace{-3mm}
 \end{figure}
For this representation, we use the spectrogram method. The spectrogram of any signal is computed using the squared magnitude of the discrete time short-time Fourier transform~(STFT)\\
\vspace{-0.25cm}
\begin{equation}
     \text{Spectrogram} (m,\omega) = \left| \sum_{k=-\infty}^{\infty}y_{\textrm{T}}[k]\text{w}[k-m]e^
     {-j\omega k} \right|^{2},
\end{equation}
where $y_{\textrm{T}}[n]$ is the pre-processed signal captured by the surveillance system, $m$ is discrete time, $\omega$ is the frequency, and $\text{w[n]}$ is a sliding window function that acts as a filter. The spectrogram analysis of the captured RF signals can reveal the transmit frequency of the signal as well as the frequency hopping patterns. The spectrogram of the signal captured from the remote controller of the DJI Phantom 4 Pro UAV (the signal in Fig.~\ref{fig:Haart_Example}) is shown in Fig.~\ref{fig:spectrogram}. In computing the spectrogram, the signal is divided into segments of length 128 with an overlap of 120 samples between adjoining segments. Then, a Hamming window is used, followed by a 256-point DFT. The spectrogram shows that the transmit frequency of the signal is 2.4~GHz.
\begin{table}[t!]
\centering
\caption{Statistical Features.}
\label{Table_2}
\begin{tabular}{|p{2.5cm}| m{3cm}|m{2cm}|m{2cm}|}
\hline
Features & Formula & Measures\\
\hline
Mean ($\mu$)& $\frac{1}{N}\sum_{i=1}^N x_{i}$ & Central tendency \\
Absolute mean ($\bar{x}$)&  $\frac{1}{N}\sum_{i=1}^N \left|{x_{i}}\right|$  & Central tendency \\
Standard deviation($\sigma_{T}$)~& $\left[\frac{1}{N-1}\sum_{i=1}^N (x_{i}-\bar{x})^2\right]^{\frac{1}{2}}$ & Dispersion \\
Skewness ($\gamma$)& $\frac{\sum_{i=1}^N (x_{i}-\bar{x})^3}{(N-1)\sigma_{T}^3}$ &Asymmetry/shape descriptor~\\
Entropy ($H$) & $-\sum_{i=1}^N x_{i}\log_{2}x_{i}$ & Uncertainty\\
Root mean square ($x_{\rm{rms}}$)~& $\left[\frac{1}{N}\sum_{i=1}^N x_{i}^2\right]^\frac{1}{2}$ & Magnitude/Average power\\
Root ($x_{\rm{r}}$) & $\left[\frac{1}{N}\sum_{i=1}^N \left|{x_{i}}\right|^\frac{1}{2}\right]^2$ &   Magnitude\\
Kurtosis ($k$) & $\frac{\sum_{i=1}^N (x_{i}-\bar{x})^4}{(N-1)\sigma_{T}^4}$ & Tail/shape descriptor~\\
Variance & $\frac{1}{N}\sum_{i=1}^N (x_{i}-\mu)^2$ & Dispersion \\
Peak value ($x_{\rm{pv}}$)& $\rm{max(x_{i})}$ & Amplitude \\
Peak to peak ($x_{\rm{ppv}}$) & $\rm{max(x_{i})}-\rm{min(x_{i}})$ & Waveform amplitude\\
Shape factor($x_{\rm{sf}}$)& $\frac{x_{\rm{rms}}}{\bar{x}}$ & Shape descriptor \\
Crest factor &  $\frac{x_{\rm{max}}}{x_{\rm{rms}}}$  & Peak extremity \\
Impulse factor & $\frac{x_{\rm{max}}}{\bar{x}}$ & Impulse\\
Clearance factor & $\frac{x_{\rm{max}}}{x_{\rm{r}}}$ & Spikiness \\
\hline
\end{tabular}
\vspace{-3mm}
\end{table}

The spectrogram displays the energy distribution of the signal along the time-frequency axis. Therefore, the energy trajectory can be computed from the spectrogram by taking the maximum energy values along the time-axis. From this distribution, we estimate the energy transient by searching for the most abrupt change in the mean or variance of the normalized energy trajectory. The energy transient defines the transient characteristics of the signal in energy domain. For the RF signal in Fig.~\ref{fig:Haart_Example}, the normalized energy trajectory and the corresponding energy transient are shown in Fig.~\ref{fig:energy_trajectory}.

Once the energy transient is detected, RF fingerprints (a set of 15 statistical features) are extracted. Each feature is a physical descriptor of the energy transients and can provide valuable information for ML-based classification of the signals captured from different UAV controllers. Table~\ref{Table_2} gives the list of the extracted features used in this study. The features extracted from 17 UAV controllers are used to train five different ML algorithms: kNN, RandF, discriminant analysis (DA), support vector machine (SVM), and neural networks (NN). Since some of the features may be correlated, therefore redundant, we also perform feature selection to reduce the computational cost of the classification algorithm.

\subsection{Feature Selection Using NCA}\label{NCA_Explanation}
The NCA algorithm is a nearest neighbor-based feature weighting algorithm, which learns a feature weighting vector by maximizing a leave-one-out classification accuracy using a gradient based optimizer. It is a non-parametric, embedded, and supervised learning method for feature selection. NCA learns the weighting vector/matrix by which the primary data are transformed into a lower-dimensional space~\cite{goldberger2005neighbourhood}. In this lower-dimensional space, the features are ranked according to a weight metric, with the more important features receiving higher weight values.

Given a set of training samples representing the different UAV controllers, $U=\{(\boldsymbol{x_{1}},Y_{1}),\dotsc,(\boldsymbol{x_{i}},Y_{i}),\dotsc,(\boldsymbol{x_{n}}, Y_{n})\}$, where $\boldsymbol{x_{i}}$ is a $p$-dimensional feature vector extracted from the energy transient, $Y_{i}\in \{1,2,\dotsc,C\}$ are the corresponding class labels, and $C$ is the number of classes. Then the NCA learns the feature weighting vector $\textbf{w}$ by maximizing a regularized objective function $f(\textbf{w})$ with respect to the weight of each features. The regularized objective function is defined as:
\begin{equation}
\begin{aligned}
\label{eq:NCA_optimization}
     f(\textbf{w})&=\frac{1}{n}\sum_{i=1}^{n}\left[\sum_{j=1,i\neq j}^{n}{p_{ij}Y_{ij}}-\lambda\sum_{r=1}^{p}{{{w}^2_r}}\right],\text{where}\\
    p_{ij}&=\begin{cases}
    \frac{k(d_{\bf{w}}(\boldsymbol{x}_{i},\boldsymbol{x}_{j}))}{\sum_{j=1,i\neq j}^{n} k(d_{\bf{w}}(\boldsymbol{x}_{i},\boldsymbol{x}_{j}))} & \text{if}~i\neq j\\
    0, & \text{if}~i=j
    \end{cases},\\
  Y_{ij}&=\begin{cases}
         1, & \text{if}~Y_{i}=Y_{j}\\
         0, & \text{otherwise}
\end{cases},
\end{aligned}
\end{equation}
$n$ is the number of samples in the feature set, $\lambda$ is the regularization term, $w_r$ is a weight associated with the $r$th feature, and $p_{ij}$ is the probability with which each point $\boldsymbol{x}_{i}$ selects another point $\boldsymbol{x}_{j}$ as its reference neighbor and inherits the class label of the latter~\cite{yang2012neighborhood}. The parameter $Y_{ij}$ is an indicator function, $d_{\textbf{w}}(\boldsymbol{x}_{i},\boldsymbol{x}_{j})=\sum_{r=1}^{p}{{{w}^2_r}}|{x}_{ir}-{x}_{jr}|$ is a weighted distance function between $\boldsymbol{x}_{i}$ and $\boldsymbol{x}_{j}$, and $k(a)=\exp(\frac{a}{\sigma})$ is some kernel function. Thus, NCA is a kernel-based feature selection algorithm that selects the most descriptive and informative features by optimizing (\ref{eq:NCA_optimization}) using gradient update techniques.



\begin{figure}[t]
 \center
 \includegraphics[trim=0.1cm 0cm 0.1cm 0.6cm, clip,width=0.47\textwidth]{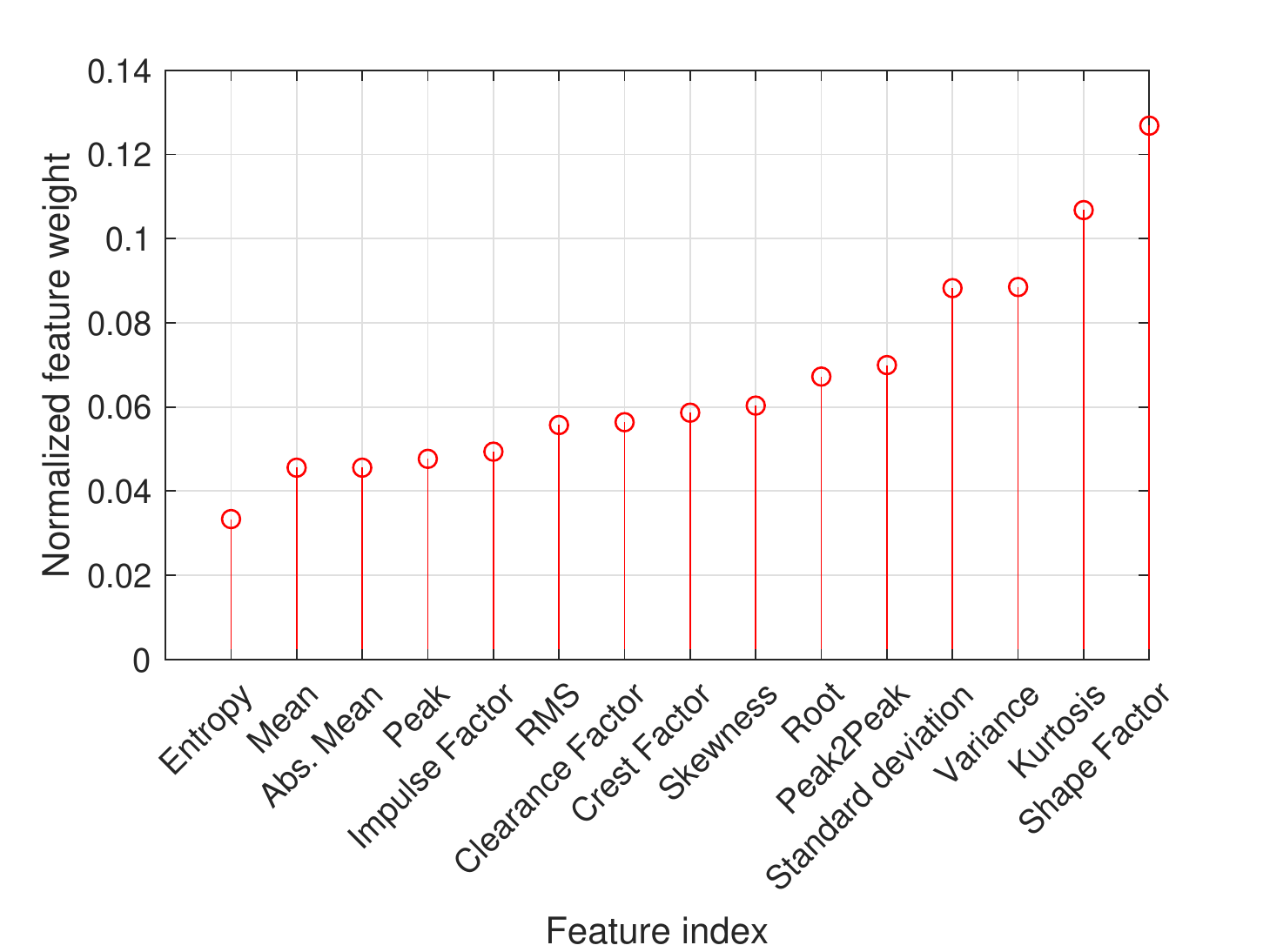}
 \caption{NCA ranking of all the 15 RF fingerprints extracted from 17 UAV controllers.}
 \label{Fig:NCA_result17cont}
 \vspace{-3mm}
\end{figure}

Fig.~\ref{Fig:NCA_result17cont} shows the results of the NCA ranking of 15 features extracted from the 17 UAV controllers. The experimental setup and structure of the captured data are described in Section~\ref{four}. In Fig.~\ref{Fig:NCA_result17cont}, we see that NCA ranks the RF fingerprints according to their weight values. It turns out that the shape factor is the most discriminative feature in the feature set. The next significant feature is the kurtosis which describes the tailedness of the energy trajectory curve. Next are variance and standard deviation. On the other hand, entropy, which measures the uncertainty in the data set, is the least significant feature.
Based on these results, the ML algorithms can safely discard the less important features and still achieve good (even better) classification performance. This is because discarding the less significant features reduces the chance of overfitting. In addition, for large-scale classification problems, there can be huge computational saving in training and testing the classifiers with fewer number of features.







\begin{table}[t!]
\centering
\caption{UAV Catalogue.}
\label{UAV_catalogue}
\begin{tabular}{|c|c|c|c|}
\hline
Make & Model & Make & Model\\
\hline
\multirow{5}{*}{\text{DJI}}&Inspire 1 Pro&\multirow{5}{*}{\text{Spektrum}}& DX5e\\
& Matrice 100 && DX6e\\
& Matrice 600$^1$ && DX6i\\
& Phantom 4 Pro$^1$ && JR X9303\\
& Phantom 3 &  &\\
\hline
Futaba & T8FG & Graupner & MC-32\\
\hline
HobbyKing & HK-T6A & FlySky & FS-T6\\
\hline
Turnigy & 9X&Jeti Duplex & DC-16\\
\hline
\end{tabular}
\vspace{-3mm}
\end{table}
\addtocounter{footnote}{1}
\footnotetext{A pair of these controllers is used in this study. For all other controllers, only one of each type is considered.}

\section{Experimental Setup}\label{four}
During the experiments, RF signals are captured from 17 UAV controllers, six mobile Bluetooth devices (smart phones), and a Wi-Fi router. Table~\ref{UAV_catalogue} gives the catalogue of the UAV controllers from eight different manufacturers. All the UAV controllers transmit control signals in the 2.4~GHz frequency band. In particular, a pair of UAV controllers from DJI Matrice 600 and DJI Phantom 4 Pro models are used while only one of the other controller type is used. This is important for forensic and security analysis to investigate the confusion that would arise when a target recognition system attempt to distinguish between UAV controllers of the same make and model. For the remaining part of the study we will refer to the pair of DJI Matrice 600 as DJI M600 Mpact and DJI M600 Ngat. Similarly, the pair of Phantom 4 Pro controllers will be referred to as DJI Phantom 4 Pro Mpact and DJI Phantom 4 Pro Ngat. \looseness=-1

 \begin{figure}{}
\center{
 \begin{subfigure}[]{\includegraphics[scale=0.55]{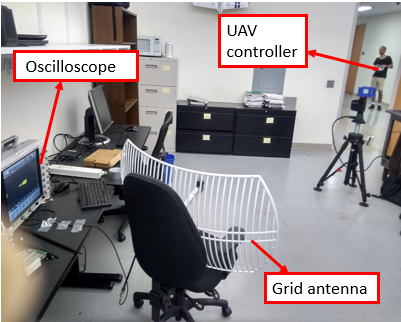}\label{fig:indoor}}
\end{subfigure}
\hfill
 \begin{subfigure}[]{\includegraphics[scale=0.55]{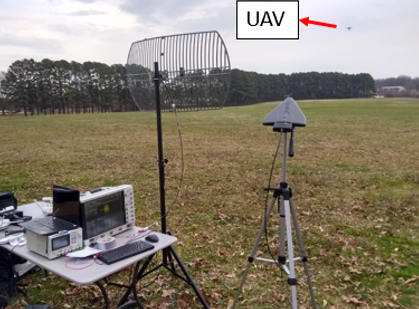}\label{fig:outdoor}}
\end{subfigure}
 \caption{{(a) Indoor and (b) outdoor experimental scenarios for UAV signal detection}.}
\label{fig:Experimental_scenario}}
\vspace{-4mm}
 \end{figure}

Fig.~\ref{fig:Experimental_scenario} shows the indoor and outdoor experimental scenarios. In each case, the RF passive surveillance system detects signals transmitted by the UAV controllers and the interference sources. Due to space limitations, only the results of the indoor experiments will be reported. The experimental RF passive surveillance system  consists of a 6~GHz bandwidth Keysight MSOS604A oscilloscope with a maximum sampling frequency of 20~Gsa/s, 2~dBi omnidirectional antenna (for short distance detection), and 24~dBi Wi-Fi grid antenna (for longer distance detection). The antennas operate in the 2.4~GHz frequency band. Detection range for the far field scenario can be further improved by using a combination of high-gain receive antennas and low noise power amplifiers (LNAs).

The receiver antenna senses the environment for the presence of signals from UAV controllers. The collected data are automatically saved in a cloud database for post processing. For each controller, 100 RF signals are collected. Each RF signal contains $5000$k samples and has a time span of 0.25~ms. The database are partitioned with the ratio $p=0.2$. That is, 80\% of the saved data is randomly selected for training and the remaining 20\% is used for testing (4:1 partitioning).

\begin{figure}[t!]
 \center
 \includegraphics[trim=1cm 0.2cm 1.2cm 0.5cm, clip,scale=0.5,width=0.95
 \linewidth]{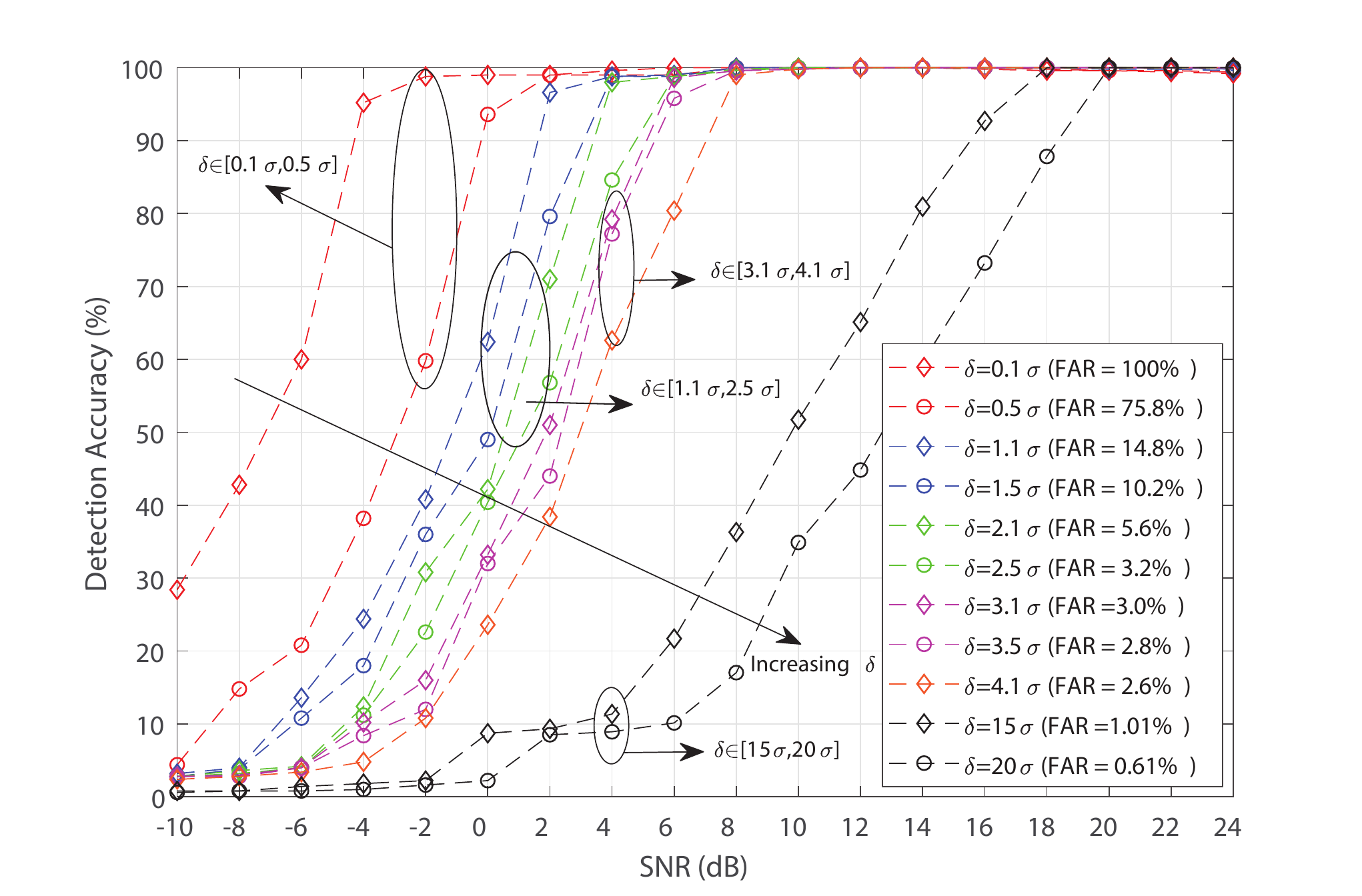}
 \caption{The signal detection accuracy of the Markov model-based na\"ive Bayesian detector versus SNR for different values of $\delta$.}
 \label{Fig:Detection_accuracy}
 \vspace{-3mm}
\end{figure}

\section{Results}\label{results_1}
\subsection{Detection Results}\label{results_detection}

Detection performance of the proposed system is assessed for different SNRs and threshold choices, and the results are presented in Fig.~\ref{Fig:Detection_accuracy}. The selected thresholds are functions of the standard deviation ($\sigma$) of the preprocessed noise data and the FAR specification. The value of $\sigma$ is estimated after performing multiresolution analysis (wavelet preprocessing) of a concatenation of several noise data captured from the environment. On the other hand, FAR, also known as the probability of false detection, is the percentage of false alarms per the number of non-events.
\begin{table*}[t!]
\centering
\begin{threeparttable}
\caption{Performance Of The ML Classification  Algorithms At 25 \si{\deci\bel} SNR. 100 Sample Signals From Each UAV Controller Is Captured With 80\% Used For Training and 20\% For Testing (Partition Ratio $=0.2$). The Selected
RF fingerprints Are: Shape Factor, Kurtosis, And Variance.}
\label{Table_classificationAccuracy}
\begin{tabular}{|c| c| c |c |c |c |c|}
\hline
$\#$ of controllers & Classifier & \multicolumn{2}{c|}{Accuracy (\%)$^2$}  & \multicolumn{2}{c|}{Computational Time (s)$^2$} \\

 & & All Feat. & Selected Feat.& All Feat.& Selected Feat.\\
 \hline
\multirow{5}{*}{15}&kNN&97.30&98.13&24.85&24.57\\
&DA&96.30&94.43&19.42&18.58\\
&SVM&96.47&91.67&119.22&111.02\\
&NN&96.73&96.13&38.73&38.14\\
&RandF&98.53&97.73&21.37&20.89\\
\hline
\multirow{5}{*}{17}&kNN&95.62&95.53&26.16&25.13\\
&DA&92.77&88.12&19.36&18.90\\
&SVM&93.82&87.88&139.94&141.68\\
&NN&92.88&93.03&46.04&43.33\\
&RandF&96.32&95.18&24.71&24.84\\
\hline
\end{tabular}
\end{threeparttable}
\end{table*}

 Fig.~\ref{Fig:Detection_accuracy} shows that at very low SNR, such as $-10$~dB, the detection accuracy is generally very low irrespective of the threshold. As a result, in case of low-level signals (where signals completely buried in the noise), the probability of missed detection increases. Besides, for a given SNR, it is observed that the set threshold also affects the performance of the detection system. For instance, when the system operates at an SNR of 2~dB, a threshold of $\delta=0.1\sigma$ will achieve a detection accuracy of above 99\%.
 However, the threshold $\delta=0.1\sigma$ yields to a FAR of 100\%. Therefore, a very low threshold value will result in a high percentage of misclassification of the noise data as signals. Furthermore, for the given SNR of 2~dB, an increase in the threshold value to $\delta=1.1\sigma$ will reduce the detection accuracy and FAR to 96.6\% and 14.8\%, respectively. Further increase in the threshold to $\delta=2.5\sigma$ will greatly reduce the detection accuracy and FAR to 40.4\% and 3.2\%, respectively. Therefore, the optimum threshold depends on the operating condition and the requirements on the FAR. Besides, the input impedance of the oscilloscope places a fundamental limit on the sensitivity of the passive detection system used in this study.

In addition, Fig.~\ref{Fig:Detection_accuracy} shows that better detection performance (with low FAR) can be achieved if the detector operates at higher SNRs (above 8~dB) and threshold $\delta\in[2.5\sigma, 4.1\sigma]$. For instance, when the receiver operates at an SNR of 10~dB with a threshold $\delta=3.5\sigma$, the detection accuracy becomes 99.8\% and, FAR drops to 2.8\%. Although a continuous increase in the threshold will further reduce the FAR, it will not always guarantee a better detection accuracy, especially when the receiver operates at SNRs of less than 18~dB.
This is because the dissimilarity between the transition matrices of the RF signal and noise classes reduces as $\delta$ increases beyond some optimum value. Therefore, there is a high chance of detection error as $\delta$ increases indefinitely.






Once a signal has been detected, the bandwidth and the modulation-based features are estimated as described in Section~\ref{second_stage_detector}. This information is used to decide if the signal comes from a UAV controller or any of the known interference sources (Wi-Fi and Bluetooth sources). Given that the detected signal comes from a UAV controller, it is sent to the ML-based classification system for accurate identification. Classification results are discussed next.



\subsection{UAV Classification Results}

For the classification problem, 15 statistical features given in Table~\ref{Table_2} are extracted. Feature selection is performed using the NCA algorithm as described in Section~\ref{NCA_Explanation}.


To validate the result efficiency of the NCA and the ML classifiers, 10 Monte Carlo simulations are run on the test dataset. On one hand, all the 15 features are used for the UAV controller classification problem. On the other hand, only three most significant features are used according to the NCA weight ranking shown in Fig.~\ref{Fig:NCA_result17cont}. These are the shape factor, kurtosis and variance. The classification experiments are run separately for the case of 15 and 17 UAV controllers. Here, the number of controllers represents the number of classes considered. In the case of 15 controllers, all the controllers are of a different model. However, in the case of 17 controllers, a pair of DJI Matrice 600 (labeled as DJI Matrice 600 Mpact and DJI  Matrice 600 Ngat) and a pair DJI Phantom 4 Pro controllers (labeled as DJI Phantom 4 Pro Mpact and DJI Phantom 4 Pro Ngat) are considered in addition to 13 different models.\looseness=-1

Table~\ref{Table_classificationAccuracy} provides the classification accuracy of all five ML algorithms. With the exception of the kNN classifier, the table shows that the classification accuracy is only slightly higher when all the features are used as compared to when only the three selected features are used. Therefore, there is little performance loss in using only the selected fingerprints. In addition, there are computational savings in time and storage when only the selected fingerprints are used for the classification. The savings in time and computational resources may be critical in aerial surveillance systems, where the response time to effectively neutralizing a threat is very small. Moreover, as the number of UAVs (to be classified) increases, the savings in time will become more substantial. Hence, the results in Table~\ref{Table_classificationAccuracy} validates the decision to perform feature selection using the NCA algorithm.



Table~\ref{Table_classificationAccuracy} shows the RandF classifier yields the highest classification accuracy when all the features are used. For the case of 15 and 17 controllers, RandF achieves an accuracy of 98.53\% and 96.32\%, respectively. Therefore, when all the features are used, RandF is the best performing classifier. It is followed by the kNN classifier, which achieves an accuracy of 97.30\% and 95.62\% with 15 and 17 controllers, respectively. The DA classifier is the least optimal when all the features are utilized. On the other hand, when only the three selected features are used, the kNN classifier performs the best with an accuracy of 98.13\% and 95.53\% for 15 and 17 controllers, respectively. It is followed by the RandF classifier which an accuracy of 97.73\% and 95.18\% with 15 and 17 controllers, respectively. When only the three most significant features are used, the least optimal classifier is SVM. We also note that the DA classifier has the shortest computational time whereas the SVM classifier has the longest computational time.

 \begin{figure}[t]
\center{
 \begin{subfigure}[]{\includegraphics[trim=0.6cm 0cm 0.0cm 1.5cm, clip,width=0.43\textwidth]{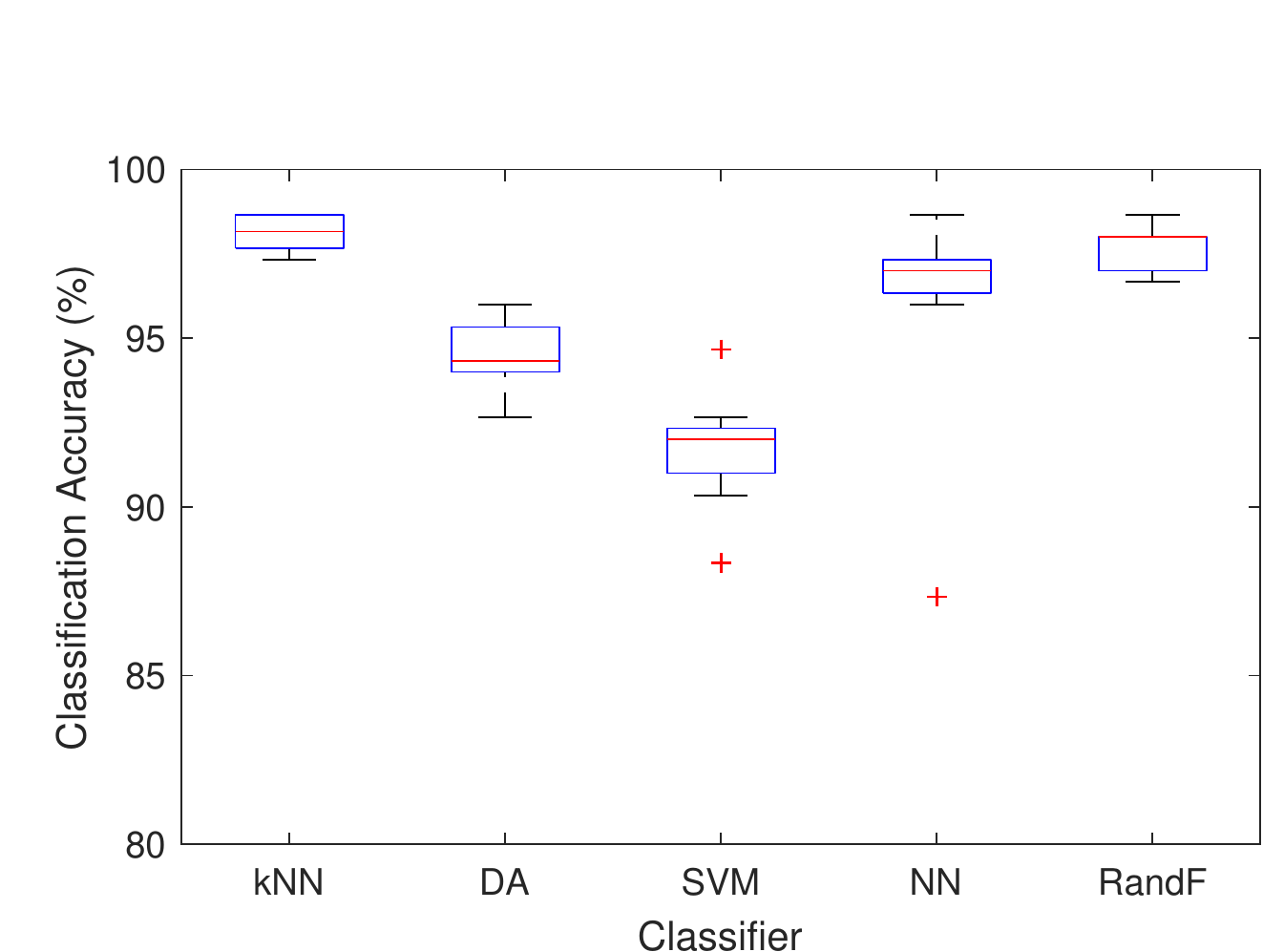}\label{fig:box_15contr}}
\end{subfigure}
\hfill
 \begin{subfigure}[]{\includegraphics[trim=0.6cm 0cm 0.0cm 1.5cm, clip,width=0.43\textwidth]{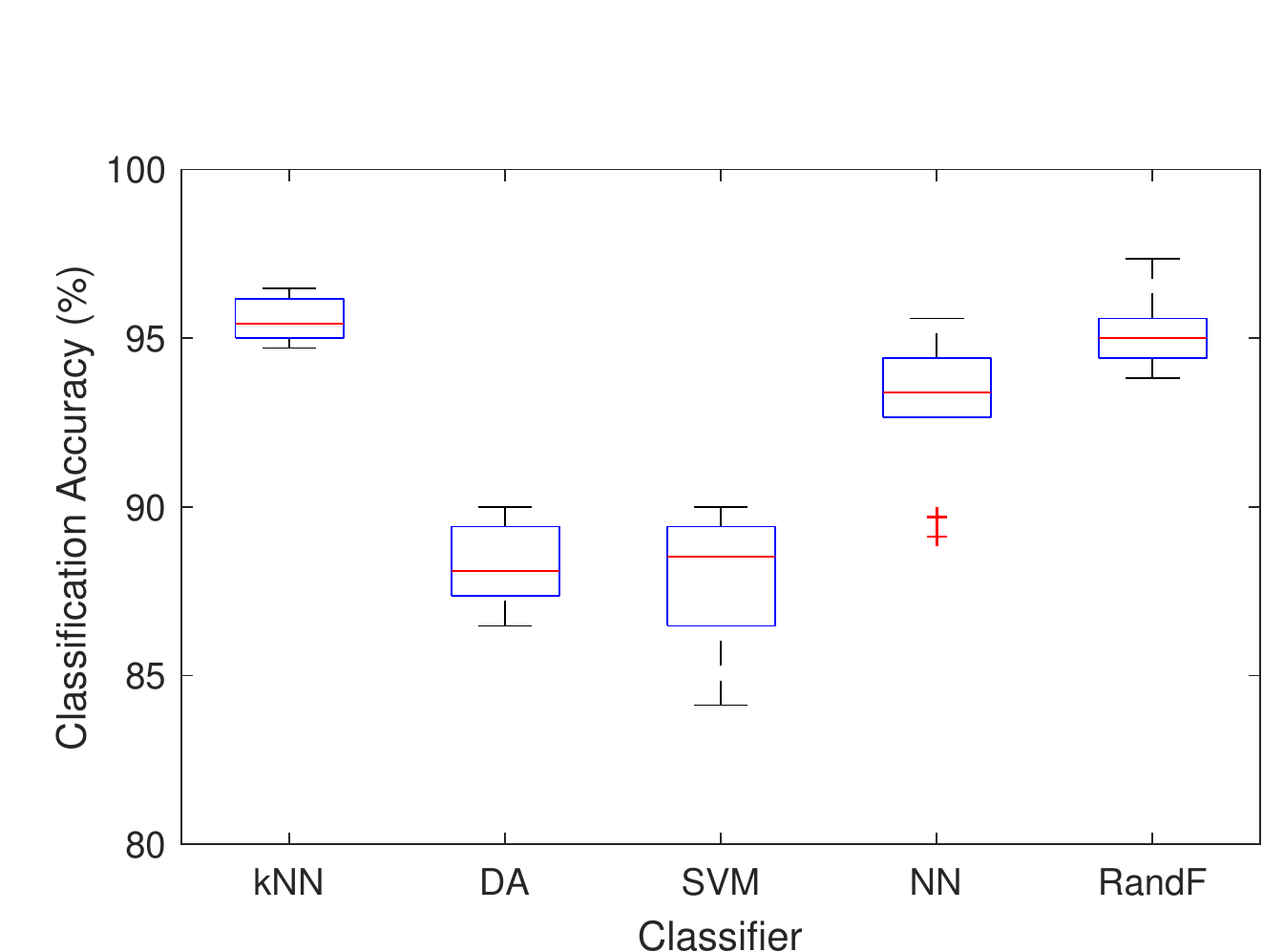}\label{fig:Box_17contr}}
\end{subfigure}
 \caption{{Box plot analysis of the classification accuracy of the ML classifiers using the three selected features (shape factor, kurtosis, and variance) with (a) 15 controllers, and (b) 17 controllers}.}
\label{fig:Box_plot_1}}
\vspace{-3mm}
 \end{figure}
 \begin{figure}[t!]
 \center
 \includegraphics[width=0.42\textwidth]{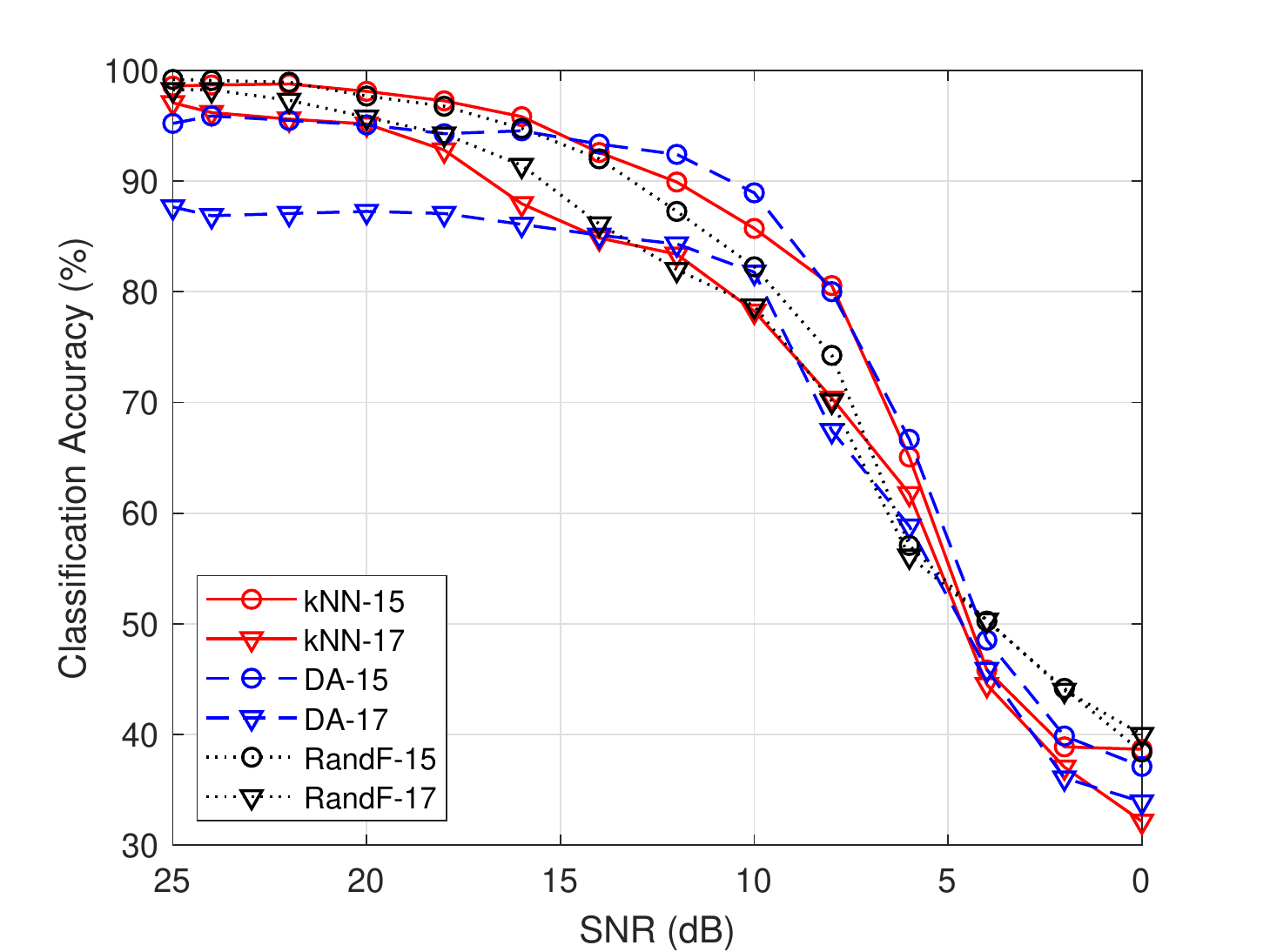}
 \caption{Classification accuracy versus SNR for kNN, RandF and DA classifiers using the three selected RF fingerprints (shape factor, kurtosis, and variance) as features for training and testing the ML classifiers.}
 \label{Fig:KNN_classification_accuracy}
 \vspace{-3mm}
\end{figure}

 \begin{figure*}{}
\center{
 \begin{subfigure}[kNN with 15 controllers]{\includegraphics[scale=0.6]{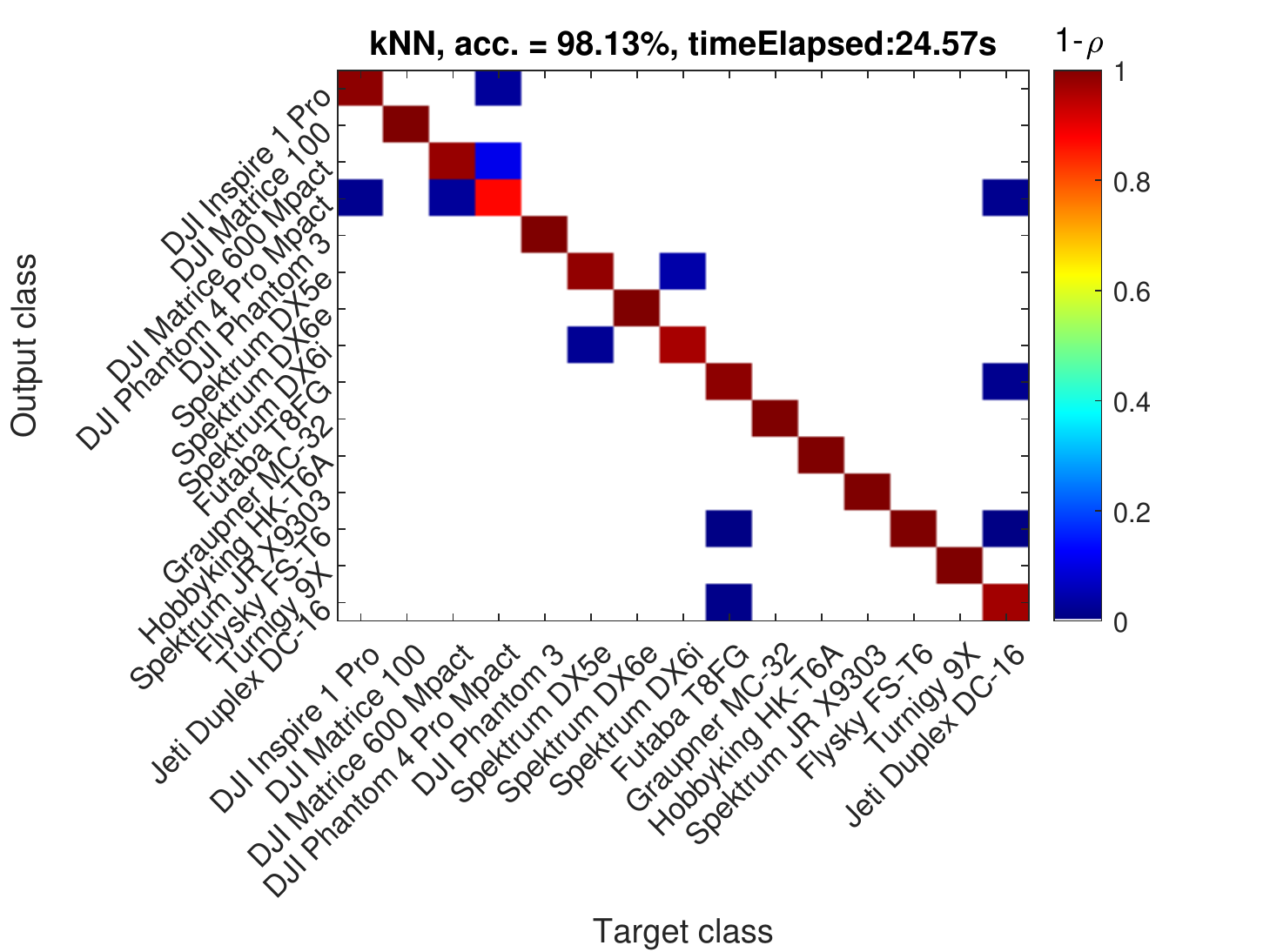}\label{fig:kNN_15}}
\end{subfigure}
\begin{subfigure}[ kNN with 17 controllers]{\includegraphics[scale=0.6]{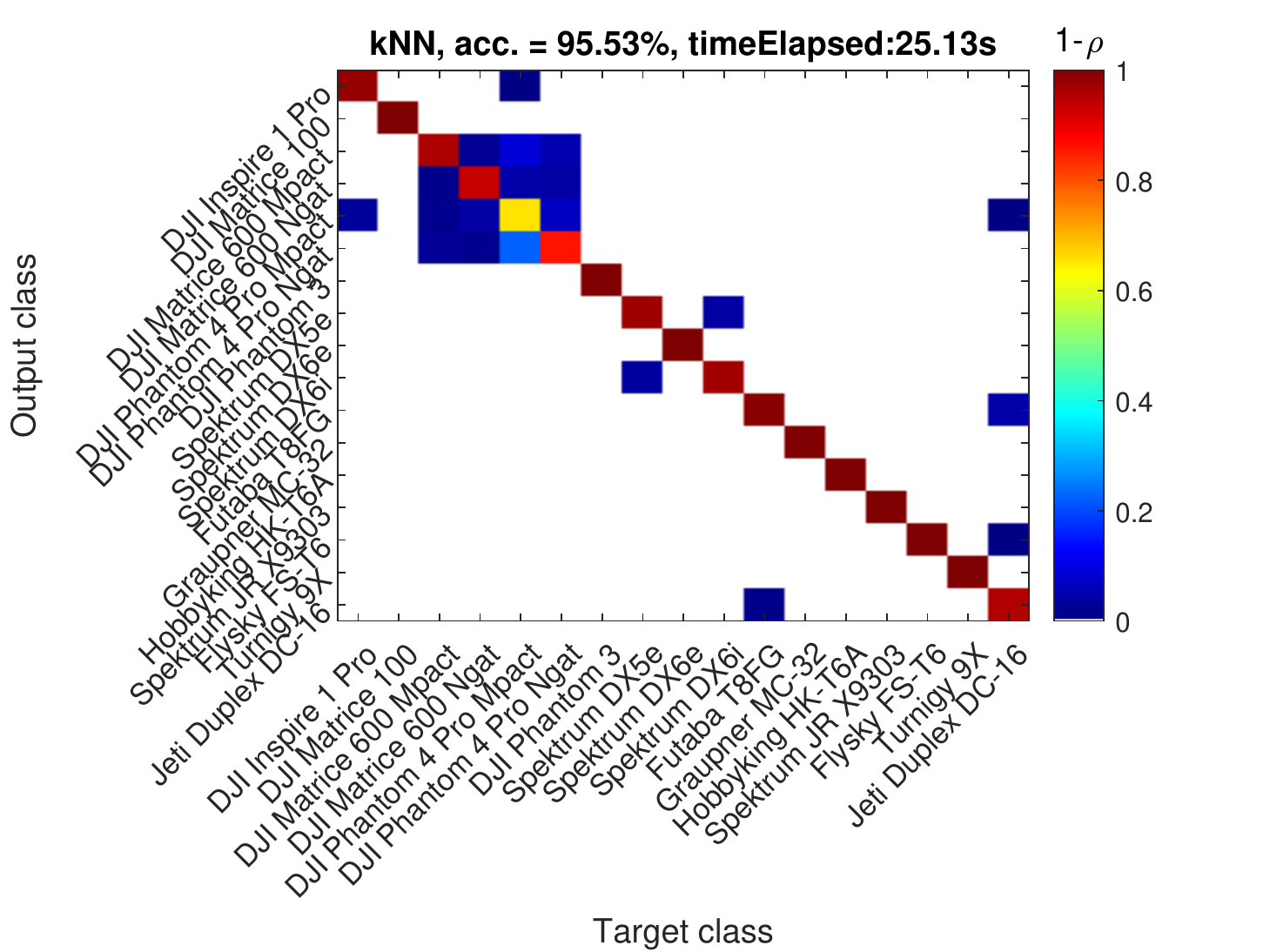}\label{fig:kNN_17}}
\end{subfigure}
\hfill
\begin{subfigure}[RandF with 15 controllers]{\includegraphics[scale=0.6]{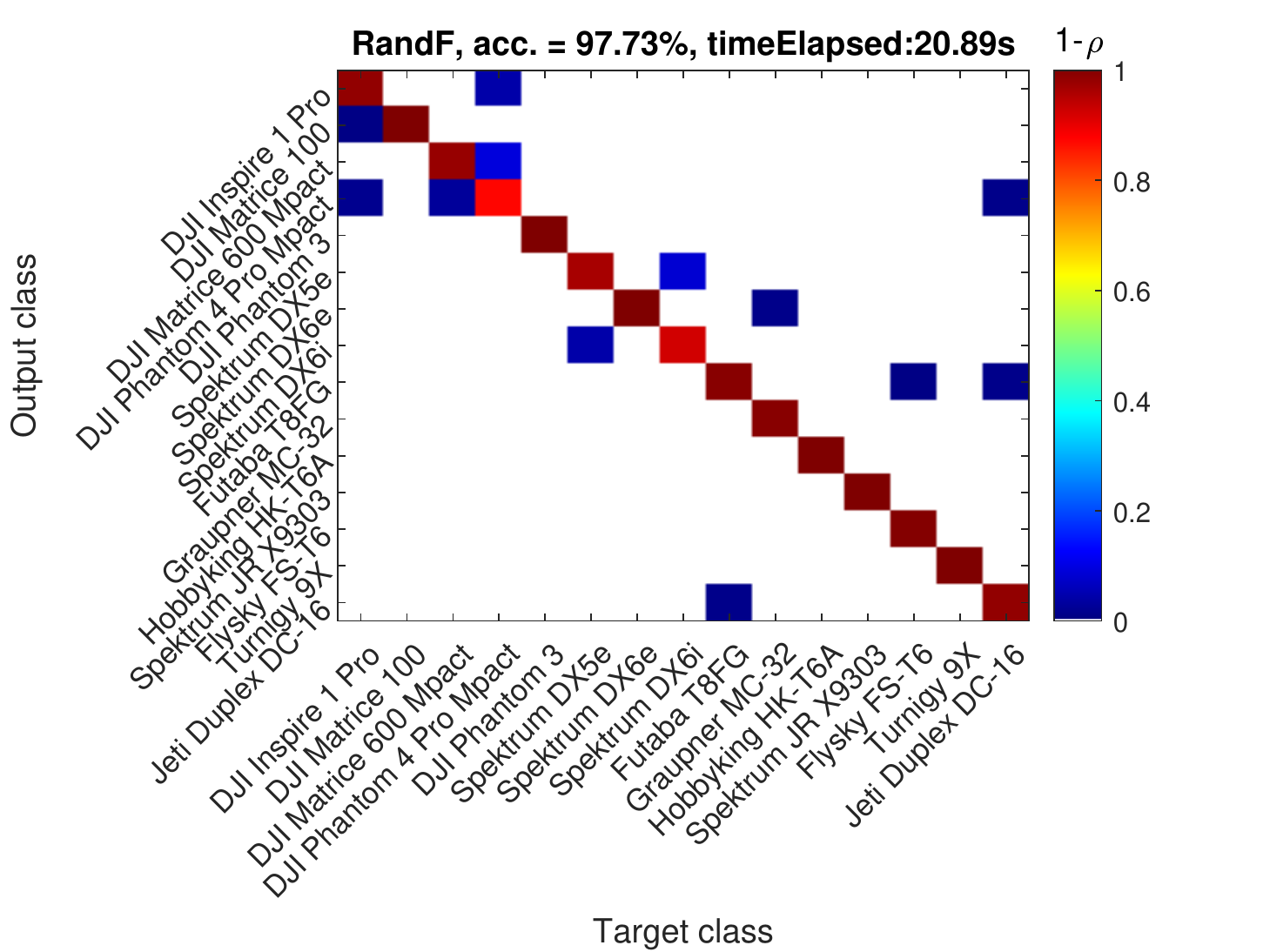}\label{fig:RF_15}}
\end{subfigure}
 \begin{subfigure}[RandF with 17 controllers]{\includegraphics[scale=0.6]{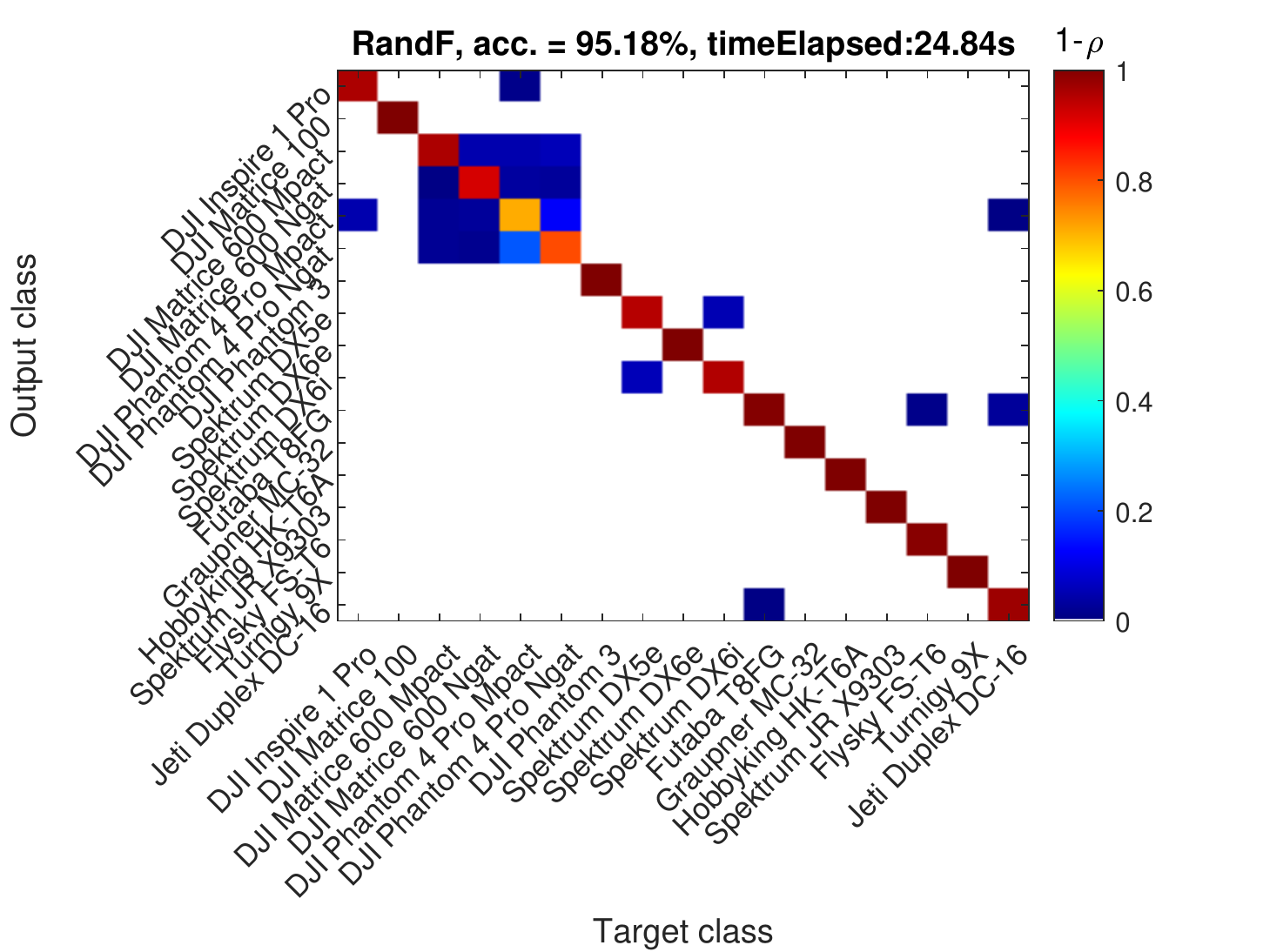}\label{fig:RF_17}}
\end{subfigure}
\hfill
\begin{subfigure}[DA with 15 controllers]{\includegraphics[scale=0.6]{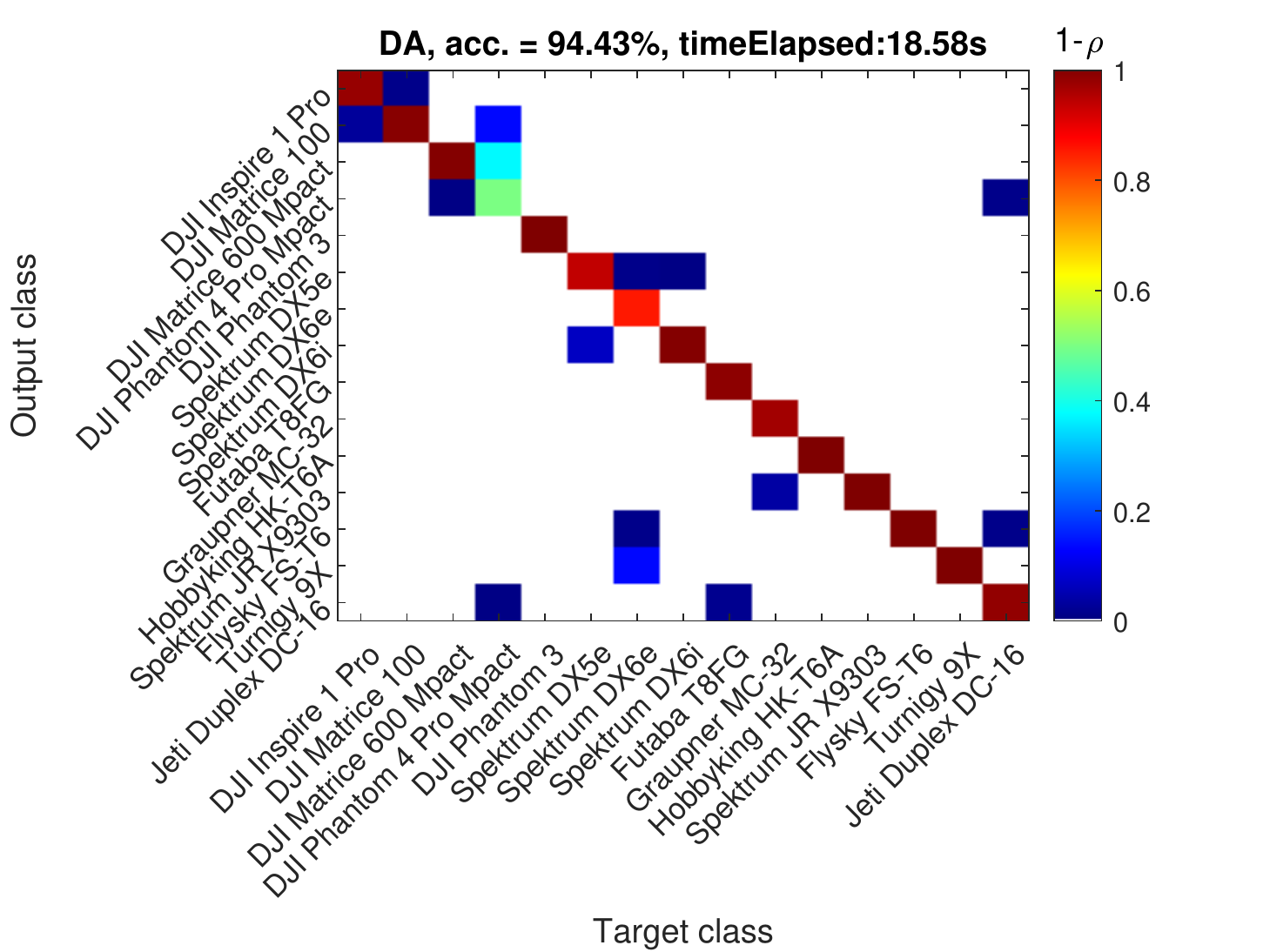}\label{fig:SVM_15}}
\end{subfigure}
\begin{subfigure}[DA with 17 controllers]{\includegraphics[scale=0.6]{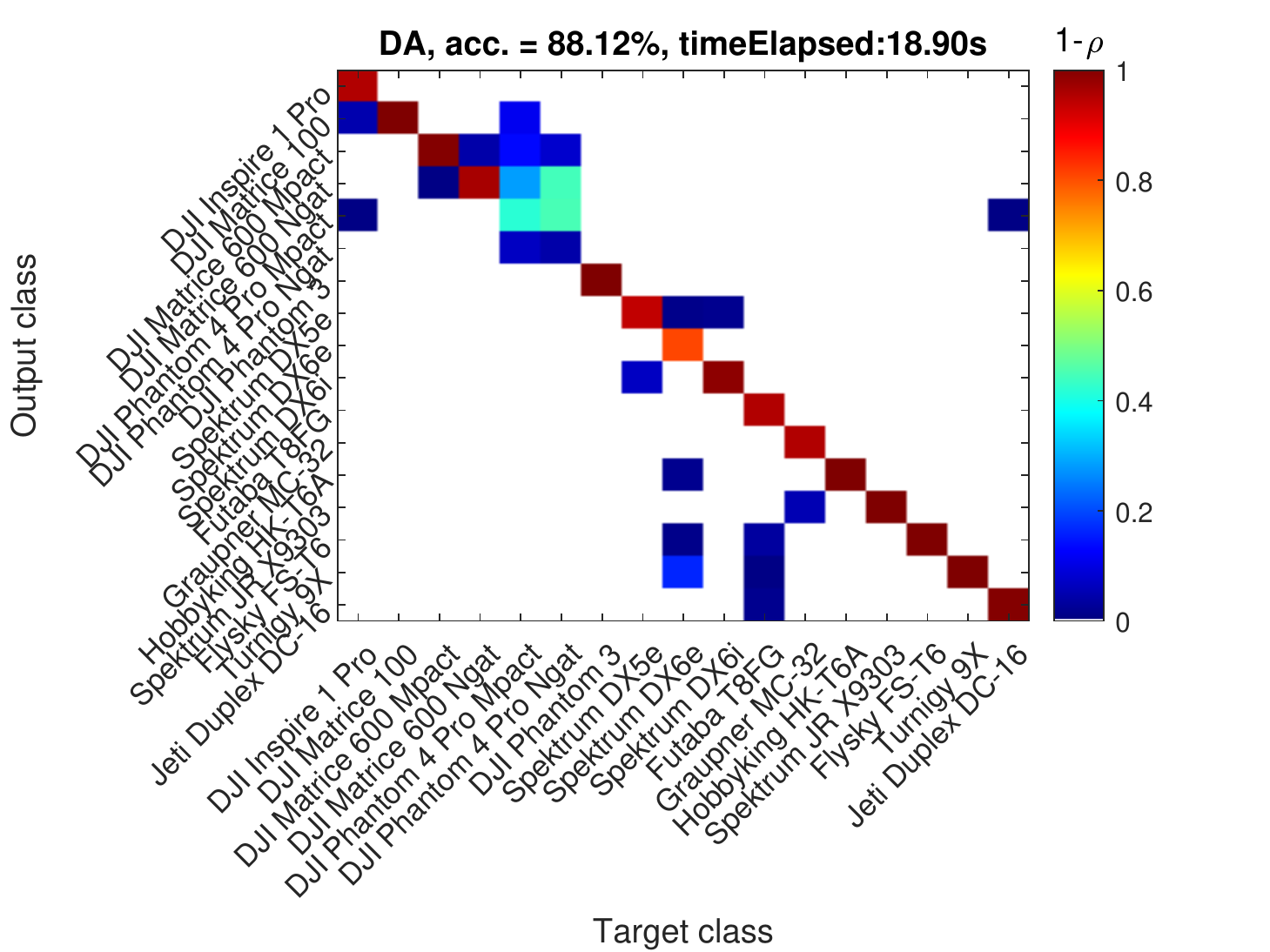}\label{fig:SVM_17}}
\end{subfigure}}

 \caption{{Confusion matrices of kNN, RandF and DA classifiers using the three selected RF fingerprints (shape factor, kurtosis, and variance). In the confusion matrices, the colorbar is used to specify the degree of confusion in terms of the confusion probability $\rho$. Moving down the colorbar, the degree of confusion increases with increasing value of $\rho$.}
\label{fig:confusion_matrix_1}}
 \end{figure*}
\addtocounter{footnote}{1}
\footnotetext{Both the accuracy and total computation time are the average of the 10 Monte Carlo simulations.}

Table~\ref{Table_classificationAccuracy} provides only the average classification accuracy results. A more detailed summary can be obtained from a box plot analysis shown in Fig.~\ref{fig:Box_plot_1}. Each box plot gives a summary of the performance of a classifier in terms of the minimum, first quartile, median (red horizontal line), third quartile, and the maximum accuracy values over 10 Monte Carlo simulations. Comparing the box plots in Fig.~\ref{fig:box_15contr} and Fig.~\ref{fig:Box_17contr}, we see that the box plot metrics for each classifier are lower in the case of 17 controllers as compared to the case of 15 controllers. This will be further investigated with the help of the confusion matrix. In addition, the box plots reveal the presence of outliers in the performance of the SVM and NN classifiers. These outliers suggest that for a given test signal, SVM and NN classifiers could produce accuracy values well below the average values reported in Table~\ref{Table_classificationAccuracy}. This observation raises the concern about the reliability of these classifiers for the UAV controller classification problem.

The SNR of the detected signal is an important factor that influences the accuracy of the classifiers.  Fig.~\ref{Fig:KNN_classification_accuracy} shows the accuracy versus SNR for the kNN, RandF and DA classifiers. For signals with SNR in the interval between 15 and 25~dB, the kNN is slightly better than the RandF for the case of 15 controllers. In the same SNR region, the RandF performs best for the case of 17 controllers. In this SNR range, the DA classifier has the worst performance. On the other hand, for SNR between 4 and 15~dB, the performance of the DA classifier improves significantly, outperforming the kNN and RandF classifiers when 15 controllers are considered. This is an interesting observation since DA is known to have the shortest computational time. However, for SNR between 0 to 4~dB, the RandF classifier has the best performance. In general, the accuracy of all the classifiers increases with SNR. Therefore, to ensure accurate identification of the UAV controller, it is best to operate the receiver at SNR above 15~dB, in which case, kNN and RandF are the optimal classifiers for the datasets. Fig.~\ref{Fig:KNN_classification_accuracy} also shows that for all SNR, the accuracy plot is slightly lower when 17 controllers are considered as compared to the case of 15 controllers.


 The confusion matrix gives an idea of what a classifier is getting right and the type of errors it makes. Fig.~\ref{fig:confusion_matrix_1} shows the confusion matrices of the  classifiers: kNN, RandF and DA for the case of 15 and 17 remote controllers. On the vertical axis of each confusion matrix is the output class or the prediction of the classifier while the horizontal is the target class or true label. From the confusion matrices in Fig.~\ref{fig:confusion_matrix_1}, we observe that in the case of 17 controllers, the degree of confusion around the DJI controllers is relatively higher as compared to the case of 15 controllers. This is because in the former, we intentionally included two pairs of identical DJI controllers (DJI Matrice 600 MPact, DJI Matrice 600 Ngat, DJI Phantom 4 Pro Mpact, and DJI Phantom 4 Pro Ngat). Consequently, there are some confusions among these four controllers leading to a slight reduction in the classification accuracy in the case of 17 controllers. However, the kNN and RandF classifiers still achieves an average accuracy of 95.53\% and 95.18\%, respectively. Therefore, these classifiers are robust in identifying UAV controllers of the same make and model. On the other hand, the DA classifier is characterized by several more confusions among different controllers which reduces its average accuracy to 88.12\% in the case of 17 remote controllers. Thus, while the kNN and RandF seem to be the best classifiers, the DA classifier still performs well for the given dataset. \looseness=-1

\section{Conclusion}\label{label:conclusion}
\label{five}
In this paper, the problem of detecting and classifying RF signals from different UAV controllers is investigated. The detection system is designed to operate in the presence of wireless interference from Wi-Fi and Bluetooth sources. These interference signals are detected using a multistage detector, which estimates the bandwidth and modulation features of the detected RF signals. Once the signal from a UAV controller is detected, it is identified using RF fingerprints along with the ML-based classification techniques. Reducing the number of required features with the help of NCA, the study shows that it is possible to achieve an accuracy of 98.13\% in classifying 15 different controllers using only three features in a kNN classifier. It is also shown that the proposed system can even classify the same make and model UAV controllers without much compromising the overall accuracy. In addition, the detection and classification performance of the proposed system is tested for a range of SNR levels. In each task, the system is shown to be safe for SNR levels of above 10~dB.
Future studies will present the detection of UAVs directly from the UAV signals in outdoor scenarios and consider the potential of sensor fusion for improved UAV detection.

\balance
\bibliography{IEEEabrv,reference}
\bibliographystyle{IEEEtran}
\end{document}